\documentclass[11pt,a4paper,english,superscriptaddress,preprint,aps,prd]{revtex4}
\usepackage{amssymb}
\usepackage{amsmath} 
\usepackage{slashed}
\usepackage{graphicx}
\usepackage{babel}
\usepackage{hyperref}
\usepackage{color}
\usepackage{comment}
\usepackage{verbatim}
\usepackage{caption}
\usepackage{subcaption}
\makeatletter\AtBeginDocument{\let\@elt\relax}\makeatother
\usepackage{color}

\begin{document}

\title{Improved effective potential of Gildener-Weinberg models}

\author{Huan Souza}
\email{huan.souza@icen.ufpa.br}
\affiliation{Faculdade de F\'isica, Universidade Federal do Par\'a, 66075-110, Bel\'em, Par\'a, Brazil.}

\author{L. H. S. Ribeiro}
\email{luis.ribeiro@icen.ufpa.br}
\affiliation{Faculdade de F\'isica, Universidade Federal do Par\'a, 66075-110, Bel\'em, Par\'a, Brazil.}

\author{A.~C.~Lehum}
\email{lehum@ufpa.br}
\affiliation{Faculdade de F\'isica, Universidade Federal do Par\'a, 66075-110, Bel\'em, Par\'a, Brazil.}

\begin{abstract}
The Gildener-Weinberg models are of particular interest in the context of extensions to the Standard Model of particle physics. These extensions may encompass a variety of theories, including double Higgs models, grand unification theories, and proposals for dark matter, among others. In order to rigorously test these models experimentally, obtaining precise results is of crucial importance. In this study, we employ the renormalization group equation and its one-loop functions to obtain a deeper understanding of the higher-loop effective potential. Our findings reveal that the radiatively generated mass of the light particle in the Gildener-Weinberg approach experiences a substantial correction. Furthermore, our results suggest that not all flat directions are equivalent and some may be preferred by nature.
\end{abstract}

\maketitle

\section{Introduction}\label{Introduction}

In 1973, Coleman and Weinberg proposed a mechanism in which radiative corrections play a central role in inducing spontaneous symmetry breaking \cite{Coleman:1973jx}. This mechanism involves computing the effective potential for a massless model containing one scalar boson and examining the vacuum properties of the resulting effective theory. Since higher-loop corrections are computationally challenging, one can resort to the renormalization group improvement technique in order to obtain these contributions \cite{McKeon:1998tr,Souza:2020hjd,Elias:2003zm,Chishtie:2005hr,Elias:2004bc,Lehum:2019msl,Meissner:2008uw,Dias:2014txa,Quinto:2014zaa}. This is achieved by organizing the effective potential as a power series of logarithms and solving the renormalization group equation through perturbative methods. As a result, one can gain insights into the higher-loop effective potential by utilizing only the one-loop renormalization group functions.

In 1976, Gildener and Weinberg proposed an extension of the Coleman-Weinberg (CW) mechanism, referred to as the Gildener-Weinberg (GW) mechanism \cite{Gildener:1976ih}. This extension generalizes the CW mechanism to models containing multiple scalar bosons. In their work, Gildener and Weinberg made the assumption that scale invariance is spontaneously broken at tree level by a flat vacuum, which imposes restrictions on the coupling constants and simplifies the structure of the effective potential. This approach is particularly noteworthy in the field of particle physics as it offers a potential solution to the hierarchy problem of the Standard Model. This idea was first introduced by Lee and Pilaftsis in their study of a two Higgs model \cite{Lee:2012jn}.

The utilization of the GW mechanism in the two Higgs extension of the Standard Model has been thoroughly explored in several studies \cite{Lee:2012jn,Hashino:2015nxa,Lane:2019dbc,Lane:2018ycs,Kannike:2021iyh}. One of the distinctive features of GW models is the phenomenon of alignment, where the couplings of all scalars to fermions and gauge bosons are identical to those found in the single Higgs model \cite{Gunion:2002zf,Weinberg:1967tq}.

In addition to the two Higgs extension, another popular extension of the Standard Model is the inclusion of dark matter particles \cite{Steele:2013fka, Endo:2015ifa, Cosme:2018nly, Arcadi:2019lka, Arcadi:2021mag}. In such models, symmetry breaking can occur in a variety of ways, as discussed in Ref. \cite{Chataignier:2018kay}. One potential avenue for symmetry breaking is through the GW mechanism. In these models, it is essential to conduct precise calculations, as the cosmological properties of dark matter are dependent on its mass and its coupling to the Higgs boson.

The present study focuses on developing a technique for calculating the leading logarithmic (LL) contributions to the effective potential using the renormalization group improvement method in GW models. We consider the two scalars model, which serves as a toy model for the two Higgs model or any other extension of the Standard Model (SM) that involves an additional scalar particle. The calculation of LL contributions can play a substantial role in phenomenological analyses.

The study is structured as follows: In Sec.~\ref{Model}, we present the two-scalar model and derive the flatness directions of the model using the GW approach. In Sec.~\ref{ll}, we compute the LL effective potential up to the forth order in the leading logarithmic expansion through the use of the renormalization group improvement technique. Finally, in Sec.~\ref{Conclusions}, we provide our conclusions.

\section{Gildener-Weinberg approach}\label{Model}

The model under investigation in this study is a system of two interacting real scalar fields, described by the Lagrangian
\begin{equation}\label{model}
\mathcal{L}=\frac{1}{2}(\partial_\mu \sigma)^2+\frac{1}{2}(\partial_\mu\phi)^2-\frac{\lambda_1}{4!}\sigma^4-\frac{\lambda_2}{4!}\phi^4-\frac{\lambda_3}{4}\sigma^2\phi^2.
\end{equation}

To analyze the effective potential as a function of both scalar fields, we utilize the GW approach \cite{Gildener:1976ih}. First, let us consider a scenario in which the tree-level potential has a nontrivial minimum along a certain direction, denoted by $\langle\rho\rangle^2=\langle\sigma\rangle^2+\langle\phi\rangle^2=\mu^2$. This implies that each field will attain a minimum value, $\langle\sigma\rangle=n_1\mu=\mu_1, \langle\phi\rangle=n_2\mu=\mu_2$, that satisfies the following constraint:
\begin{equation}
\mu_1^2+\mu_2^2=\mu^2.
\end{equation}

One way of expressing the condition $\mu_1^2 + \mu_2^2 = \mu^2$ is to represent $n_1$ and $n_2$ as the sine and cosine of an angle, respectively,
\begin{equation}
\mu_1^2 + \mu_2^2 = n_1^2\mu^2 + n_2^2\mu^2 = (\sin\alpha)^2\mu^2 + (\cos\alpha)^2\mu^2 = \mu^2.
\end{equation}

To demonstrate this in practice, the fields $\sigma$ and $\phi$ are relocated as $\sigma\to\eta_1+\sigma$ and $\phi\to\eta_2+\phi$, where $\eta_1$ and $\eta_2$ are now considered as the quantum fields, and $\sigma$ and $\phi$ serve as background fields. The Lagrangian can be expressed as
\begin{eqnarray}  \mathcal{L}&=&\frac{1}{2}(\partial_\mu\eta_1)^2+\frac{1}{2}(\partial_\mu\eta_2)^2-\frac{\lambda_1}{4!}\eta_1^4-\frac{\lambda_2}{4!}\eta_2^4-\frac{\lambda_3}{4}\eta_1^2\eta_2^2-\eta_1\Bigr(\frac{\lambda_1}{6}\sigma^3+\frac{\lambda_3}{2}\sigma\phi^2\Bigr)\nonumber\\  && -\eta_2\Bigr(\frac{\lambda_2}{6}\phi^3+\frac{\lambda_3}{2}\sigma^2\phi\Bigr)-\eta_2^2\Bigr(\frac{\lambda_1}{4}\sigma^2+\frac{\lambda_3}{4}\phi^2\Bigr)-\eta_2^2\Bigr(\frac{\lambda_2}{4}\phi^2+\frac{\lambda_3}{4}\sigma^2\Bigr)\nonumber \\  && -\lambda_3\sigma\phi\eta_1\eta_2 - \frac{\lambda_1}{6}\sigma\eta_1^3-\frac{\lambda_2}{6}\phi\eta_2^3-\frac{\lambda_3}{2}\Bigr(\sigma\eta_1\eta_2^2+\phi\eta_2\eta_1^2\Bigr).\end{eqnarray}

Based on the Lagrangian described previously, the tadpole conditions are given by
\begin{subequations}\label{tadpoles}  \begin{eqnarray}   &&\langle\eta_1(0)\rangle=-i\Bigr(\frac{\lambda_1}{6}\mu_1^3+\frac{\lambda_3}{2}\mu_1\mu_2^2\Bigr)=0; \\   &&\langle\eta_2(0)\rangle=-i\Bigr(\frac{\lambda_2}{6}\mu_2^3+\frac{\lambda_3}{2}\mu_1^2\mu_2\Bigr)=0.  \end{eqnarray} \end{subequations}

Taking into account the assumption that $\sin\alpha \neq 0$ and $\cos \alpha\neq0$ (a discussion about the cases in which $\sin\alpha = 0$ or $\cos\alpha = 0$ can be found in Appendix \ref{apxB}), we find
\begin{subequations}\label{cond}  \begin{eqnarray}   &&\lambda_1 = -3 \lambda_3 \cot ^2(\alpha );\\   &&\lambda_2 = -3 \lambda_3 \tan ^2(\alpha ).  \end{eqnarray} \end{subequations}

Assuming that $\lambda_3$ is negative, the potential is guaranteed to have a lower bound. Using Eqs.(\ref{cond}), the mass matrix can then be determined as
\begin{equation}
 M = \begin{pmatrix}
  -\lambda_3 \mu ^2 \cos ^2(\alpha ) & \lambda_3 \mu ^2 \sin (\alpha ) \cos (\alpha )\\ 
  \lambda_3 \mu ^2 \sin (\alpha ) \cos (\alpha ) & -\lambda_3 \mu ^2 \sin ^2(\alpha )
\end{pmatrix},
\end{equation}
which has the eigenvalues
\begin{subequations}
 \begin{eqnarray}
  &&m_1^2 = -\lambda_3\mu^2=|\lambda_3|\mu^2;\\
  &&m_2^2 = 0.
 \end{eqnarray}
\end{subequations}

Accordingly, we observe the generation of mass for one particle, while the other remains massless. This second particle was referred to as the ``scalon'' by Gildener and Weinberg. Although mass is generated, it is evident that the minimum of the potential is zero. This can be seen explicitly in the expression
\begin{equation}
V(\sigma,\phi) = -\frac{1}{8} \lambda_3 \sigma ^4 \cot ^2(\alpha )-\frac{1}{8} \lambda_3 \phi ^4 \tan ^2(\alpha )+\frac{1}{4} \lambda_3 \sigma ^2 \phi ^2,
\end{equation}
at its minimum,
\begin{equation}
V(\sigma,\phi)\Bigr{|}_{\substack{\sigma=\mu\sin\alpha\\\phi=\mu\cos\alpha}} = 0.
\end{equation}

\begin{figure}[h]
     \centering
     \begin{subfigure}[b]{0.3\textwidth}
         \centering
         \includegraphics[width=\textwidth]{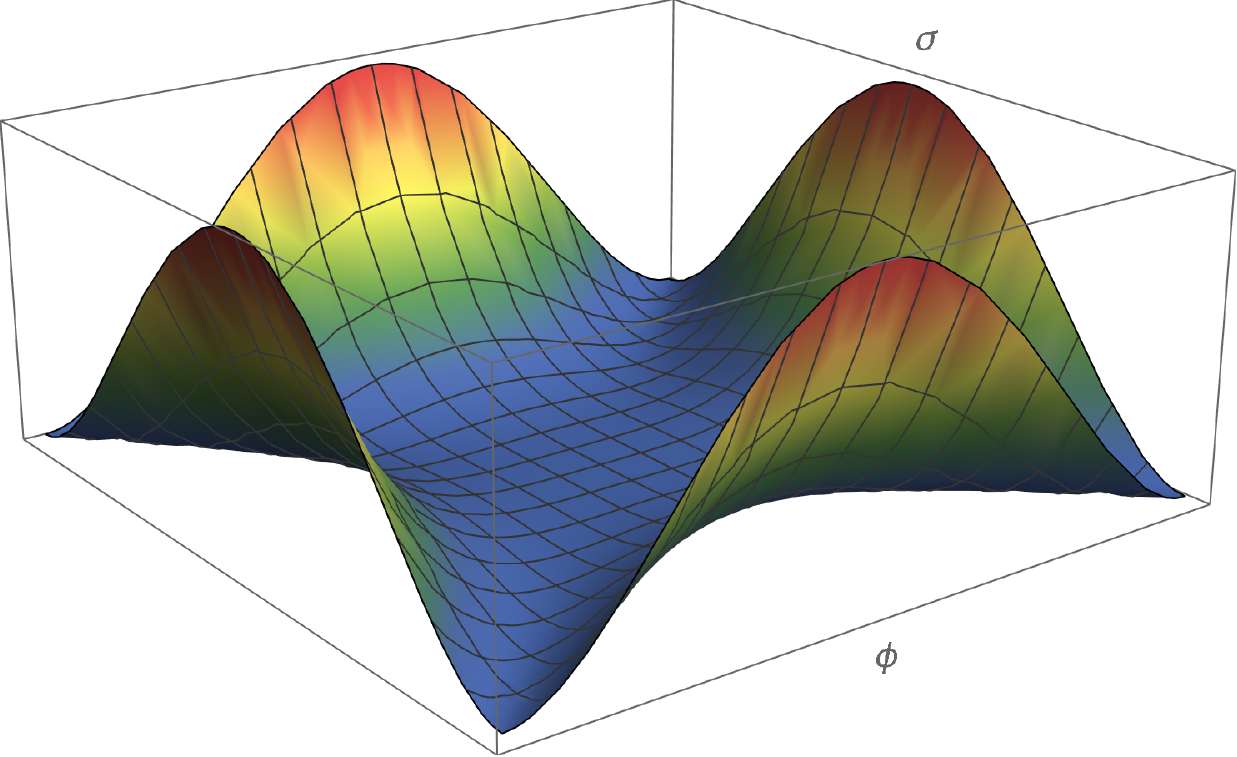}
         \caption{$\alpha=\frac{\pi}{4}$}
         \label{fig:y equals x}
     \end{subfigure}
     \hfill
     \begin{subfigure}[b]{0.3\textwidth}
         \centering
         \includegraphics[width=\textwidth]{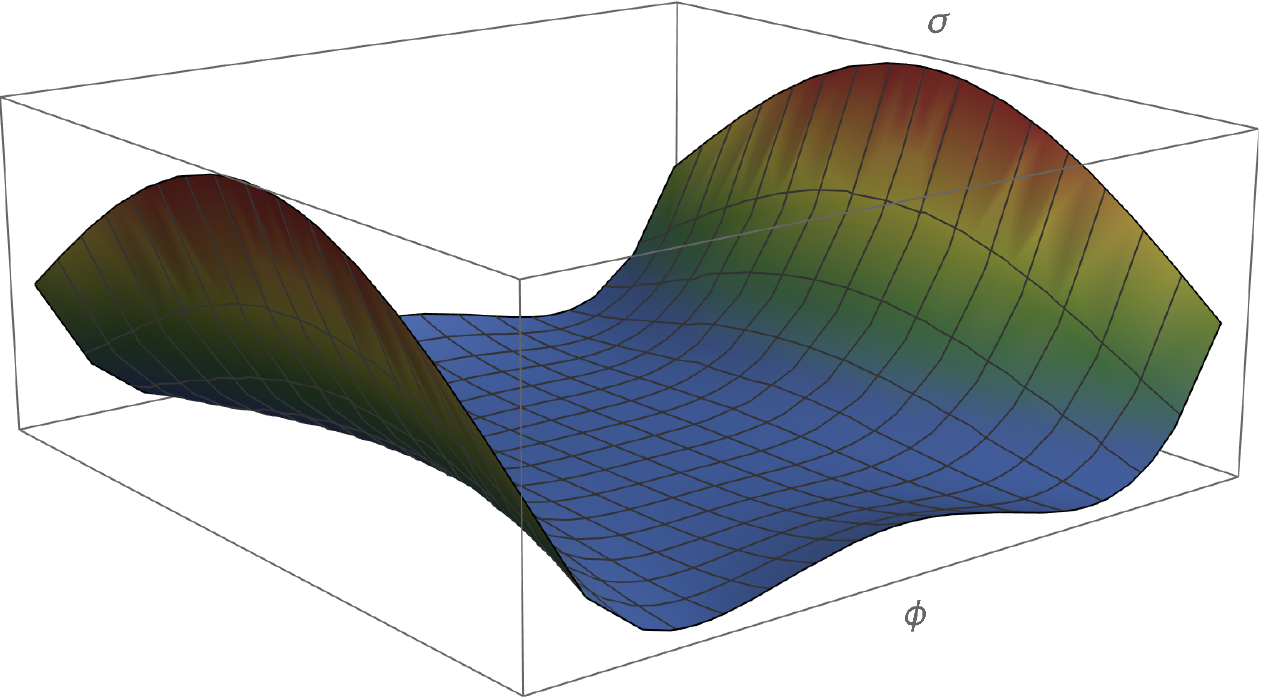}
         \caption{$\alpha=\frac{\pi}{3}$}
         \label{fig:three sin x}
     \end{subfigure}
     \hfill
     \begin{subfigure}[b]{0.3\textwidth}
         \centering
         \includegraphics[width=\textwidth]{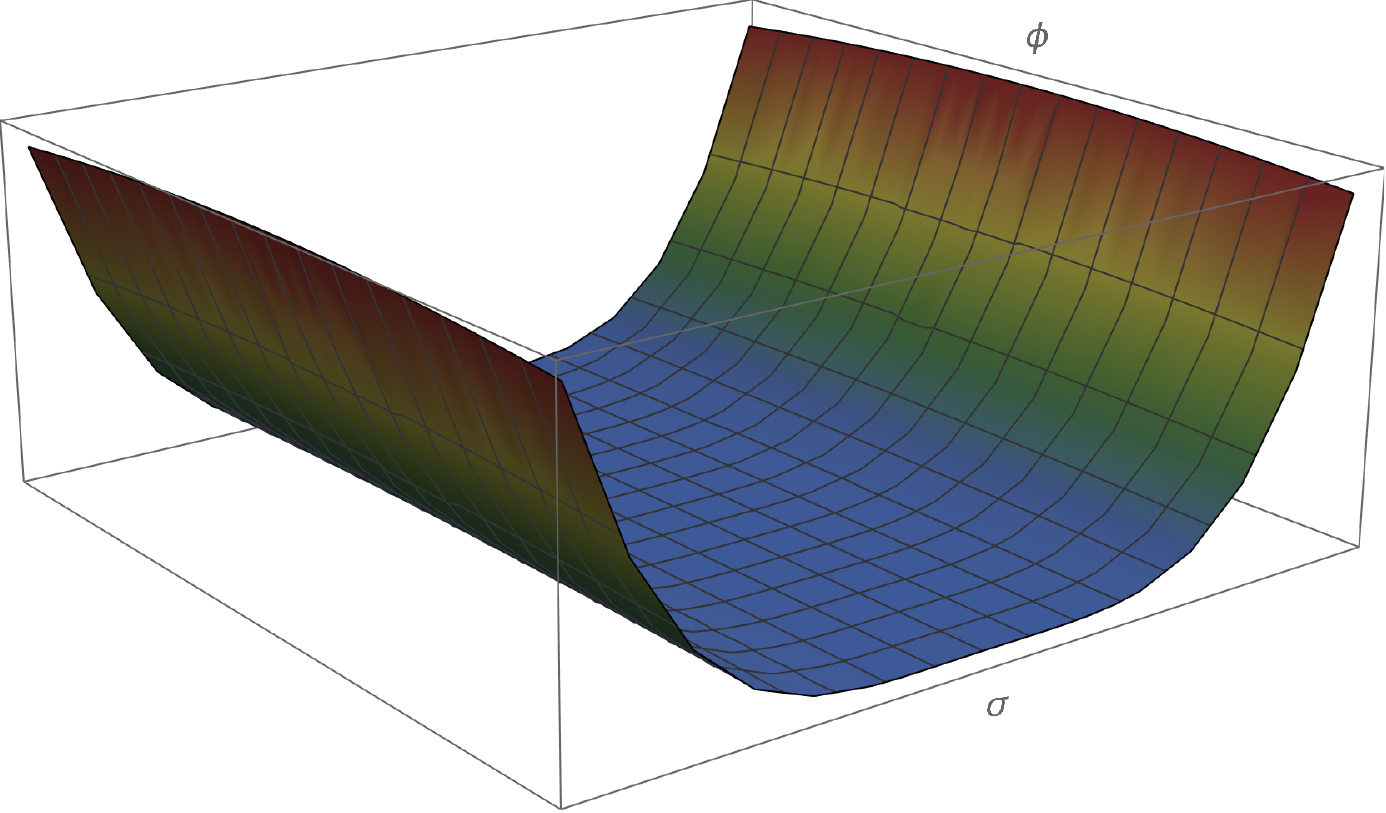}
         \caption{$\alpha=\frac{\pi}{20}$}
         \label{fig:five over x}
     \end{subfigure}
        \caption{The tree level potential acquires a flat direction.}\label{figtree}
\end{figure}

The appearance of a flat direction in the tree-level potential results in the generation of mass. This can be illustrated graphically by fixing the value of $\lambda_3=-0.1$ and comparing the results for different values of $\alpha$ (as shown in Fig.~\ref{figtree}).

It is noteworthy that, owing to $\lambda_3<0$, there exist both attractive and repulsive forces in our system, which are responsible for the mass generation at tree level.

To complete the analysis at tree level, we determine the eigenvectors of the mass matrix and the diagonalization matrix. The eigenvectors are expressed as
\begin{subequations}
 \begin{eqnarray}
  \vec{u}_1=\begin{pmatrix}
   -\cot(\alpha) & 1
  \end{pmatrix};
  \end{eqnarray}
  \begin{equation}
   \vec{u}_2 = \begin{pmatrix}
                  \tan(\alpha) & 1
                 \end{pmatrix},
  \end{equation}
\end{subequations}
and the diagonalization matrix is given by
\begin{equation}
 S = \left(
\begin{array}{cc}
 -\cot (\alpha ) & \tan (\alpha ) \\
 1 & 1 \\
\end{array}
\right).
\end{equation}

\section{Effective potential}\label{ll}

Having analyzed the GW tree-level potential, we can now proceed to compute its radiative corrections. These corrections are expected to shift the minimum of the potential to a nonzero value and introduce a local maximum at $\sigma=\phi=0$. The computation of the effective potential will be performed using the renormalization group improvement method, as previously done in \cite{Souza:2020hjd}.

\subsection{Summing logarithms}

The renormalization group equation (RGE) applied to the effective potential is
\begin{equation}
 \Bigr(\mu\frac{\partial}{\partial\mu}+\beta_{\lambda_1}\frac{\partial}{\partial \lambda_1}+\beta_{\lambda_2}\frac{\partial}{\partial\lambda_2}+\beta_{\lambda_3}\frac{\partial}{\partial\lambda_3}+\sigma\gamma_\sigma\frac{\partial}{\partial\sigma}+\phi\gamma_\phi\frac{\partial}{\partial\phi}\Bigr)V_{eff}=0,
\end{equation}

\noindent that we can rewrite as
\begin{equation}\label{RGE}
 \Bigr(-2\frac{\partial}{\partial L}+\beta_{\lambda_1}\frac{\partial}{\partial \lambda_1}+\beta_{\lambda_2}\frac{\partial}{\partial\lambda_2}+\beta_{\lambda_3}\frac{\partial}{\partial\lambda_3}\Bigr)V_{eff}=0,
\end{equation}

\noindent where we used the fact that the anomalous dimensions functions $\gamma_\sigma$ and $\gamma_\phi$ are vanishing (see for instance Appendix \ref{apxA}) and
\begin{eqnarray}\label{RGE1}
 \mu\frac{\partial V_{eff}}{\partial\mu}=\mu\frac{\partial V_{eff}}{\partial L}\frac{\partial L}{\partial \mu}=-2\frac{\partial V_{eff}}{\partial L}.
\end{eqnarray}

The effective potential can be computed by utilizing the RGE organizing the potential as a logarithmic power series, represented as
\begin{eqnarray}
 V_{eff}&=&\frac{\delta_1}{4!}\sigma^4+\frac{\delta_2}{4!}\phi^4+\frac{\delta_3}{4}\sigma^2\phi^2+V^{LL}+V^{NLL}+...\nonumber\\
 &=&\frac{\delta_1}{4!}\sigma^4+\frac{\delta_2}{4!}\phi^4+\frac{\delta_3}{4}\sigma^2\phi^2+\sum_{n=0}^\infty C^{LL}_nL^n+\sum_{n=1}^\infty C^{NLL}_nL^n+...
\end{eqnarray}
where $V^{LL}$ and $V^{NLL}$ are the LL and next-to-LL contributions, respectively. The coefficients of the LL series, $C^{LL}_n$, are proportional to $\xi^4 x^{n+1}$, where $\xi^4$ is a combination of the fields, such as $\sigma^2\phi^2$, $\sigma^4$, $\phi^4$, etc., and $x$ is a combination of the coupling constants, e.g. $x^2$ could be $\lambda_1^2$, $\lambda_2^2$, $\lambda_3^2$, $\lambda_1\lambda_2$, etc. The coefficients of the next-to-LL series, $C^{NLL}_n$, are proportional to $\xi^4 x^{n+2}$.

In this work, our objective is to determine the LL contributions to the effective potential. To do so, we consider only the LL series, expressed as
\begin{equation}\label{LL}
 V_{eff}=\sum_{n=0}^\infty C_n^{LL}L^n+\frac{\delta_1}{4!}\sigma^4+\frac{\delta_2}{4!}\phi^4+\frac{\delta_3}{4}\sigma^2\phi^2.
\end{equation}

Substituting the ansatz Eq.(\ref{LL}) into the RGE Eq.(\ref{RGE1}) and upon organizing the result in ascending order of the logarithmic expansion, we can identify a recurrence relation that allows us to determine all the coefficients by using the following equation:
\begin{equation}\label{recurrence}
 C_{n}^{LL}=\left(\beta_1\frac{\partial}{\partial \lambda_1}+\beta_2\frac{\partial}{\partial \lambda_2}+\beta_3\frac{\partial}{\partial \lambda_3}\right)\frac{C_{n-1}^{LL}}{2n}.
\end{equation}

\noindent With this result at hand, we are able to compute the LL contribution to the effective potential up to the desired order.

\subsection{One-loop effective potential}

The calculation of the one-loop effective potential will be carried out first. It should be noted that only values of $\alpha$ that are not equal to $\frac{n\pi}{2}$, where $n$ is an integer, will be considered. By utilizing the recurrence relation specified in Eq.\eqref{recurrence}, the one-loop effective potential can be cast as
\begin{eqnarray}\label{veffLL1l}
V^{LL}_{1}&=&\frac{(\lambda_1+\delta_1)}{24}\sigma ^4+\frac{(\lambda_2+\delta_2)}{24}\phi^4+\frac{(\lambda_3+\delta_3)}{4} \sigma^2 \phi^2+C_1^{LL}\ln \left(\frac{\sigma ^2+\phi ^2}{\mu ^2}\right),
\end{eqnarray}
\noindent where the coefficient $C_1^{LL}$ is given by
\begin{eqnarray}\label{c1LL}
C_1^{LL}=\frac{1}{256 \pi ^2}\left[ \left(\lambda_1^2+\lambda_3^2\right)\sigma^4+2 (\lambda_1+\lambda_2+4 \lambda_3)\sigma^2 \phi^2+\left(\lambda_2^2+\lambda_3^2\right)\phi^4\right].
\end{eqnarray}

We then proceed to the renormalization of the model. The renormalization conditions are described in \cite{Gildener:1976ih} and are given by Eqs. \eqref{ren_conds}, which state that
\begin{subequations}\label{ren_conds}
\begin{eqnarray}
 && \frac{\partial^4V^{LL}_1}{\partial\sigma^4}\Bigr |_{\substack{\sigma=\mu\sin\alpha\\\phi=\mu\cos\alpha}}=\lambda_1;\\
 && \frac{\partial^4V^{LL}_1}{\partial\phi^4}\Bigr |_{\substack{\sigma=\mu\sin\alpha\\\phi=\mu\cos\alpha}}=\lambda_2;\\
 && \frac{\partial^4V^{LL}_1}{\partial\sigma^2\partial\phi^2}\Bigr |_{\substack{\sigma=\mu\sin\alpha\\\phi=\mu\cos\alpha}}=\lambda_3.
\end{eqnarray}
\end{subequations}

\noindent In these conditions, the minimum of the potential is defined as the renormalization scale, $v=\mu$, and we have chosen the values $\sigma=\mu\sin\alpha$ and $\phi=\mu\cos\alpha$ at this minimum. Additionally, the condition $v_1^2+v_2^2=\mu^2$ is used to determine the counterterms as follows:
\begin{subequations}\label{1loop-ct}
\begin{eqnarray}
 &&\delta_1 = \frac{1}{1024 \pi ^2}\Bigr(2 \cos (4 \alpha ) \left(33 \lambda_1^2-34 \lambda_3 (\lambda_1+\lambda_2)+9 \lambda_2^2-94 \lambda_3^2\right)+4 \cos (6 \alpha ) (5 \lambda_1^2\nonumber\\
 &&\qquad\quad-8 \lambda_3 (\lambda_1+\lambda_2)+3 \lambda_2^2-24 \lambda_3^2)+3 \cos (8 \alpha ) \left(\lambda_1^2-2 \lambda_3 (\lambda_1+\lambda_2)+\lambda_2^2-6 \lambda_3^2\right)\nonumber\\
 &&\qquad\quad+4 \cos (2 \alpha ) \left(45 \lambda_1^2-16 \lambda_1 \lambda_3+3 \lambda_2^2-16 \lambda_3 (\lambda_2+\lambda_3)\right)-269 \lambda_1^2-22 \lambda_3 (\lambda_1+\lambda_2)\nonumber\\
 &&\qquad\quad+3 \lambda_2^2-354 \lambda_3^2\Bigr);\\
 &&\delta_2 = \frac{1}{1024 \pi ^2}\Bigr(2 \cos (4 \alpha ) \left(9 \lambda_1^2-34 \lambda_3 (\lambda_1+\lambda_2)+33 \lambda_2^2-94 \lambda_3^2\right)+4 \cos (6 \alpha ) (-3 \lambda_1^2\nonumber\\
 &&\qquad\quad+8 \lambda_3 (\lambda_1+\lambda_2)-5 \lambda_2^2+24 \lambda_3^2)+3 \cos (8 \alpha ) \left(\lambda_1^2-2 \lambda_3 (\lambda_1+\lambda_2)+\lambda_2^2-6 \lambda_3^2\right)\nonumber\\
 &&\qquad\quad+4 \cos (2 \alpha ) \left(-3 \lambda_1^2+16 \lambda_1 \lambda_3-45 \lambda_2^2+16 \lambda_3 (\lambda_2+\lambda_3)\right)+3 \lambda_1^2-22 \lambda_3 (\lambda_1+\lambda_2)\nonumber\\
 &&\qquad\quad-269 \lambda_2^2-354 \lambda_3^2\Bigr);\\
 && \delta_3 = -\frac{1}{1024 \pi ^2}\Bigr(\cos (4 \alpha ) \left(-6 \left(\lambda_1^2+\lambda_2^2\right)+28 \lambda_3 (\lambda_1+\lambda_2)+100 \lambda_3^2\right)+3 \cos (8 \alpha ) (\lambda_1^2\nonumber\\
 &&\qquad\quad-2 \lambda_3 (\lambda_1+\lambda_2)+\lambda_2^2-6 \lambda_3^2)+12 \cos (2 \alpha ) \left(\lambda_2^2-\lambda_1^2\right)+4 \cos (6 \alpha ) (\lambda_1-\lambda_2) (\lambda_1+\lambda_2)\nonumber\\
 &&\qquad\quad+11 \left(\lambda_1^2+\lambda_2^2\right)+74 \lambda_3 (\lambda_1+\lambda_2)+318 \lambda_3^2\Bigr).
\end{eqnarray}
\end{subequations}

Substituting Eq.~\eqref{1loop-ct} into Eq.~\eqref{veffLL1l}, the one-loop effective potential becomes
\begin{eqnarray}\label{renormalized_1loop}
 V^{LL}_1 &=& \frac{1}{24576 \pi ^2}\Bigr(2 \cos (4 \alpha ) (\sigma ^4 \left(33 \lambda_1^2-34 \lambda_3 (\lambda_1+\lambda_2)+9 \lambda_2^2-94 \lambda_3^2\right)+6 \sigma ^2 \phi ^2 (3 \left(\lambda_1^2+\lambda_2^2\right)\nonumber\\
 &&-14 \lambda_3 (\lambda_1+\lambda_2)-50 \lambda_3^2)+\phi ^4 \left(9 \lambda_1^2-34 \lambda_3 (\lambda_1+\lambda_2)+33 \lambda_2^2-94 \lambda_3^2\right))\nonumber\\
 &&+4 \cos (6 \alpha ) (\sigma ^4 \left(5 \lambda_1^2-8 \lambda_3 (\lambda_1+\lambda_2)+3 \lambda_2^2-24 \lambda_3^2\right)+\phi ^4 \left(-3 \lambda_1^2+8 \lambda_3 (\lambda_1+\lambda_2)-5 \lambda_2^2+24 \lambda_3^2\right)\nonumber\\
 &&-6 \sigma ^2 \phi ^2 (\lambda_1-\lambda_2) (\lambda_1+\lambda_2))+3 \cos (8 \alpha ) \left(\sigma ^4-6 \sigma ^2 \phi ^2+\phi ^4\right) \left(\lambda_1^2-2 \lambda_3 (\lambda_1+\lambda_2)+\lambda_2^2-6 \lambda_3^2\right)\nonumber\\
 &&+4 \cos (2 \alpha ) (\sigma ^4 \left(45 \lambda_1^2-16 \lambda_1 \lambda_3+3 \lambda_2^2-16 \lambda_3 (\lambda_2+\lambda_3)\right)+\phi ^4 (-3 \lambda_1^2+16 \lambda_1 \lambda_3-45 \lambda_2^2\nonumber\\
 &&+16 \lambda_3 (\lambda_2+\lambda_3))+18 \sigma ^2 \phi ^2 (\lambda_1-\lambda_2) (\lambda_1+\lambda_2))-\left(\sigma ^4 \left(269 \lambda_1^2+22 \lambda_3 (\lambda_1+\lambda_2)-3 \lambda_2^2+354 \lambda_3^2\right)\right)\nonumber\\
 &&-6 \sigma ^2 \phi ^2 \left(11 \left(\lambda_1^2+\lambda_2^2\right)+74 \lambda_3 (\lambda_1+\lambda_2)+318 \lambda_3^2\right)+\phi ^4 \left(3 \lambda_1^2-22 \lambda_3 (\lambda_1+\lambda_2)-269 \lambda_2^2-354 \lambda_3^2\right)\nonumber\\
 &&+1024 \pi ^2 \left(\lambda_1\sigma ^4+\lambda_2 \phi ^4+6 \lambda_3 \sigma ^2 \phi ^2\right)\Bigr)+\frac{1}{256 \pi ^2} \Bigr(\sigma ^4 \left(\lambda_1^2+\lambda_3^2\right)+2 \lambda_3 \sigma ^2 \phi ^2 (\lambda_1+\lambda_2+4 \lambda_3)\nonumber\\
 &&+\phi ^4 \left(\lambda_2^2+\lambda_3^2\right)\Bigr)\ln \left(\frac{\sigma ^2+\phi ^2}{\mu ^2}\right).
\end{eqnarray}

The effective potential has a dependence on the angle $\alpha$, which indicates that its form is influenced by the manner in which the spontaneous breaking of scale invariance occurs at tree level. To determine the relationships between $\lambda_1$,  $\lambda_2$, and $\lambda_3$, we apply the conditions that extremize $V^{LL}_1$,
\begin{subequations}\label{minimum}
 \begin{eqnarray}
  &&\frac{\partial V^{LL}_1}{\partial \sigma}\Bigr{|}_{\substack{\sigma=\mu\sin\alpha\\\phi=\mu\cos\alpha}} = 0;\\
  &&\frac{\partial V^{LL}_1}{\partial \phi}\Bigr |_{\substack{\sigma=\mu\sin\alpha\\\phi=\mu\cos\alpha}} = 0,
 \end{eqnarray}
\end{subequations}
\noindent finding
\begin{subequations}\label{couplingRelations}
 \begin{eqnarray}
  && \lambda_1 = -3 \lambda_3 \cot ^2(\alpha )-\frac{\lambda_3^2\csc ^4(\alpha )}{128 \pi ^2} (-37 \cos (2 \alpha )+284 \cos (4 \alpha )+29 \cos (6 \alpha )\nonumber\\
  &&\qquad\quad+68 \cos (8 \alpha )+88);\\
  && \lambda_2 = -3 \lambda_3 \tan ^2(\alpha ) -\frac{\lambda_3^2\sec ^4(\alpha )}{128 \pi ^2} (37 \cos (2 \alpha )+284 \cos (4 \alpha )-29 \cos (6 \alpha )\nonumber\\
  &&\qquad\quad+68 \cos (8 \alpha )+88).
 \end{eqnarray}
\end{subequations}

As expected from a perturbative computation, our results are given by the tree-level results plus corrections. Therefore, the effective potential reads
\begin{eqnarray}
 V_1^{LL}&=&\frac{1}{8} \lambda_3 \left(\sigma ^4 \left(-\csc ^2(\alpha )\right)-\phi ^4 \sec ^2(\alpha )+\left(\sigma ^2+\phi ^2\right)^2\right)+\frac{\lambda_3^2}{128 \pi ^2}\Bigr(136 \cos (2 \alpha ) \left(\sigma ^4-\phi ^4\right)\nonumber\\
 &&+34 \cos (4 \alpha ) \left(\sigma ^2+\phi ^2\right)^2+18 (\sigma ^4 \csc ^4(\alpha )-2 \sigma ^2 \csc ^2(\alpha ) \left(4 \sigma ^2+\phi ^2\right)-2 \phi ^2 \sec ^2(\alpha ) \left(\sigma ^2+4 \phi ^2\right)\nonumber\\
 &&+\phi ^4 \sec ^4(\alpha ))+247 \sigma ^4+222 \sigma ^2 \phi ^2+247 \phi ^4\Bigr) + \frac{\lambda_3^2}{256 \pi ^2} \Bigr(9 \sigma ^4 \csc ^4(\alpha )-6 \sigma ^2 \csc ^2(\alpha ) \left(3 \sigma ^2+\phi ^2\right)\nonumber\\
 &&-6 \phi ^2 \sec ^2(\alpha ) \left(\sigma ^2+3 \phi ^2\right)+9 \phi ^4 \sec ^4(\alpha )+10 \left(\sigma ^2+\phi ^2\right)^2\Bigr)\ln \left(\frac{\sigma ^2+\phi ^2}{\mu ^2}\right).
\end{eqnarray}

The minimum of the effective potential occurs at $(\sigma,\phi)=(\mu\sin\alpha,\mu\cos\alpha)$ if the discriminant of the mass matrix (Hessian matrix) is positive and the second partial derivative of $V^{LL}_1$ with respect to $\sigma$ is greater than zero at $(\sigma,\phi)=(\mu\sin\alpha,\mu\cos\alpha)$. The mass matrix can be expressed as follows:
\begin{equation}\label{massmatrix}
 M = \begin{pmatrix}
  \frac{\partial^2 V^{LL}_1}{\partial \sigma^2}\Bigr |_{\substack{\sigma=\mu\sin\alpha\\\phi=\mu\cos\alpha}} & \frac{\partial^2 V^{LL}_1}{\partial \sigma\partial\phi}\Bigr |_{\substack{\sigma=\mu\sin\alpha\\\phi=\mu\cos\alpha}}\\ 
  \frac{\partial^2 V^{LL}_1}{\partial \phi\partial\sigma}\Bigr |_{\substack{\sigma=\mu\sin\alpha\\\phi=\mu\cos\alpha}} & \frac{\partial^2 V^{LL}_1}{\partial \phi^2}\Bigr |_{\substack{\sigma=\mu\sin\alpha\\\phi=\mu\cos\alpha}}
\end{pmatrix},
\end{equation}
in which
\begin{subequations}
 \begin{eqnarray}
  &&\frac{\partial^2 V^{LL}_1}{\partial \sigma^2}\Bigr |_{\substack{\sigma=\mu\sin\alpha\\\phi=\mu\cos\alpha}} = -\frac{\lambda_3 \mu ^2 \csc ^2(\alpha )}{128 \pi ^2} (-16 \lambda_3 \cos (2 \alpha )+17 \lambda_3 \cos (8 \alpha )+\left(86 \lambda_3-16 \pi ^2\right) \cos (4 \alpha )\nonumber\\
  &&\qquad\qquad\qquad\qquad+57 \lambda_3+16 \pi ^2);\\
  &&\frac{{\partial^2 V^{LL}_1}}{\partial \phi^2}\Bigr |_{\substack{\sigma=\mu\sin\alpha\\\phi=\mu\cos\alpha}} = -\frac{\lambda_3 \mu ^2 \sec ^2(\alpha )}{128 \pi ^2} (16 \lambda_3 \cos (2 \alpha )+17 \lambda_3 \cos (8 \alpha )+\left(86 \lambda_3-16 \pi ^2\right) \cos (4 \alpha ) \nonumber\\
  &&\qquad\qquad\qquad\qquad+57 \lambda_3+16 \pi ^2);\\
  && \frac{\partial^2 V^{LL}_1}{\partial \sigma\partial\phi}\Bigr |_{\substack{\sigma=\mu\sin\alpha\\\phi=\mu\cos\alpha}} = \frac{\lambda_3 \mu ^2 \csc (\alpha ) \sec (\alpha )}{128 \pi ^2} (17 \lambda_3 \cos (8 \alpha )+\left(86 \lambda_3-16 \pi ^2\right) \cos (4 \alpha )+65 \lambda_3\nonumber\\
  &&\qquad\qquad\qquad\qquad+16 \pi ^2);\\
  && \frac{\partial^2 V^{LL}_1}{\partial \phi\partial\sigma}\Bigr |_{\substack{\sigma=\mu\sin\alpha\\\phi=\mu\cos\alpha}} = \frac{\lambda_3 \mu ^2 \csc (\alpha ) \sec (\alpha )}{128 \pi ^2} (17 \lambda_3 \cos (8 \alpha )+\left(86 \lambda_3-16 \pi ^2\right) \cos (4 \alpha )+65 \lambda_3\nonumber\\
  &&\qquad\qquad\qquad\qquad+16 \pi ^2),
 \end{eqnarray}
\end{subequations}
where its discriminant is $|M|=-\frac{\lambda_3^3\mu^4}{8\pi^2}+\mathcal{O}(\lambda_3^4)=\frac{|\lambda_3|^3\mu^4}{8\pi^2}+\mathcal{O}(\lambda_3^4)>0$, confirming that $(\sigma,\phi)=(\mu\sin\alpha,\mu\cos\alpha)$ represents a minimum for the effective potential. 

The masses of the physical states $h_1$ and $h_2$ are determined by the eigenvalues of the matrix $M$, that are given by
\begin{subequations}
 \begin{eqnarray}
  m_{1}^2 &=& |\lambda_3| \mu ^2-\frac{\lambda_3^2 \mu ^2 (76 \cos (4 \alpha )+17 (\cos (8 \alpha )+3)) \csc ^2(2 \alpha )}{32 \pi ^2};\\
 m_{2}^2 &=& \frac{\lambda_3^2 \mu ^2}{8 \pi ^2}.
 \end{eqnarray}
\end{subequations}

It can be observed that the mass $m_1^2$ comprises the tree-level mass plus one-loop corrections, whereas the second physical state acquires a nonzero mass due to radiative corrections. Furthermore, it is noteworthy that the mass $m_1^2$ is dependent on the angle $\alpha$, whereas $m_2^2$ is not. This $\alpha$ dependence of $m_1^2$ may result in undesirable outcomes for certain values of $\alpha$. By selecting a value of $\alpha$ that is close to, for example, $\frac{\pi}{2}$, it is possible to obtain a negative value for $m_1^2$ (see for instance Fig.~\ref{mass_alpha}). As we shall see in the next subsection, this issue arises as a problem of the one-loop approximation. In higher order calculations there are some values of $\alpha$ that represent the true minimum of the effective potential, thus eliminating the problem {Another potential concern is the occurrence of large logarithms, given that our perturbative parameter relies on $\lambda_3^{n+1}L^n$. To ensure the validity of our analyses, we must limit our analysis to values close to the minimum, as otherwise, the large logarithms could invalidate our results. We provide additional details on this matter in Appendix \ref{LargeLogs}.}

Additionally, the eigenvectors and the diagonalization matrix of the mass matrix are given by
\begin{subequations}
 \begin{eqnarray}
  \vec{u}_1 = \begin{pmatrix}
   -\cot(\alpha) + \frac{3}{16 \pi ^2} \cos (2 \alpha ) \csc ^3(\alpha ) \sec (\alpha ) &\hspace{0.5cm} 1
  \end{pmatrix};
 \end{eqnarray}
 \begin{equation}
  \vec{u}_2 = \begin{pmatrix}
   \tan(\alpha) + \frac{3}{16 \pi ^2} \cos (2 \alpha ) \csc(\alpha ) \sec^3 (\alpha ) &\hspace{0.5cm} 1
  \end{pmatrix},
 \end{equation}
\end{subequations}
and
\begin{equation}
 S_1 = \left(
\begin{array}{cc}
 -\cot (\alpha )+\frac{3 \cos (2 \alpha ) \csc ^3(\alpha ) \sec (\alpha )}{16 \pi ^2} &\hspace{0.5cm} \tan (\alpha )+\frac{3 \cos (2 \alpha ) \csc (\alpha ) \sec ^3(\alpha )}{16 \pi ^2} \\
 1 & 1 \\
\end{array}
\right).
\end{equation}

\begin{figure}[h]
     \centering
     \begin{subfigure}[b]{0.3\textwidth}
         \centering
         \includegraphics[width=\textwidth]{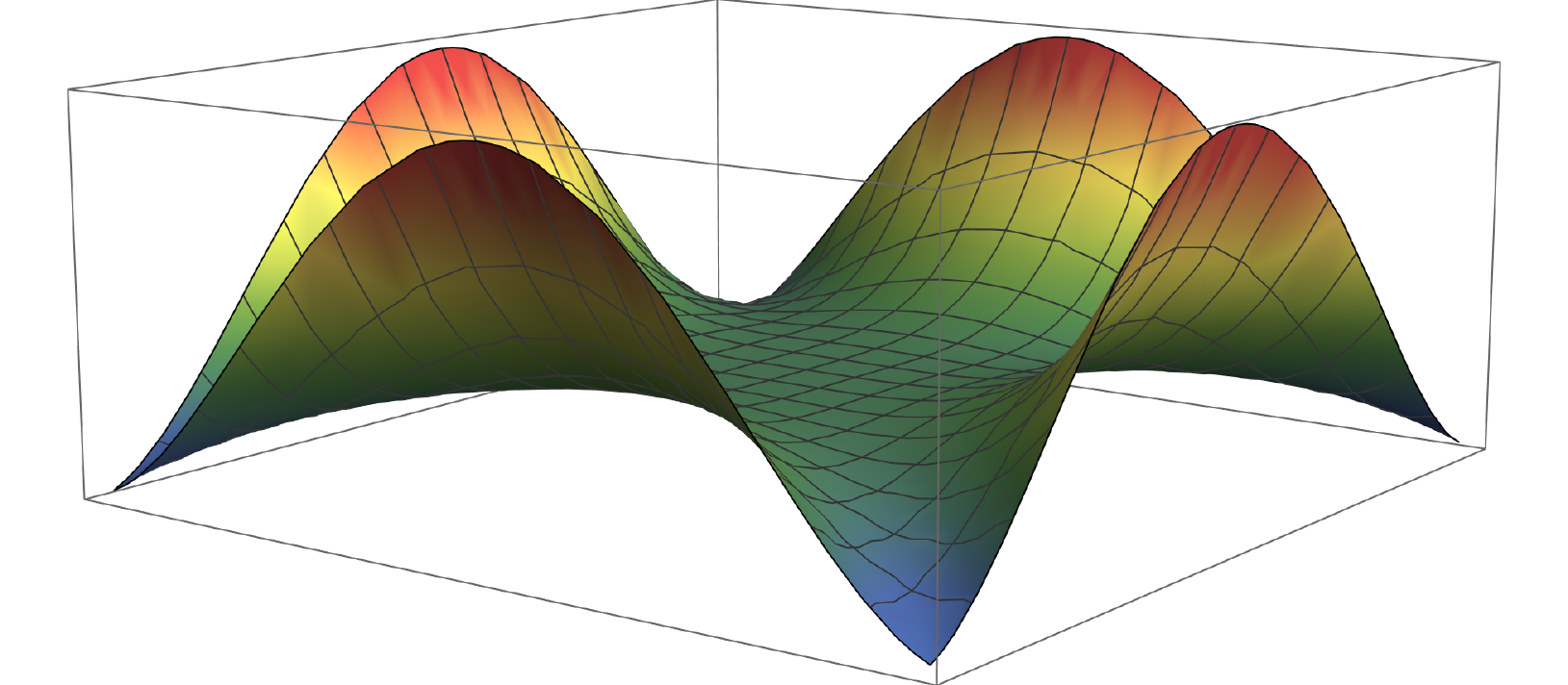}
         \caption{$\alpha=\frac{\pi}{4}$}
         \label{fig:y equals x}
     \end{subfigure}
     \hfill
     \begin{subfigure}[b]{0.3\textwidth}
         \centering
         \includegraphics[width=\textwidth]{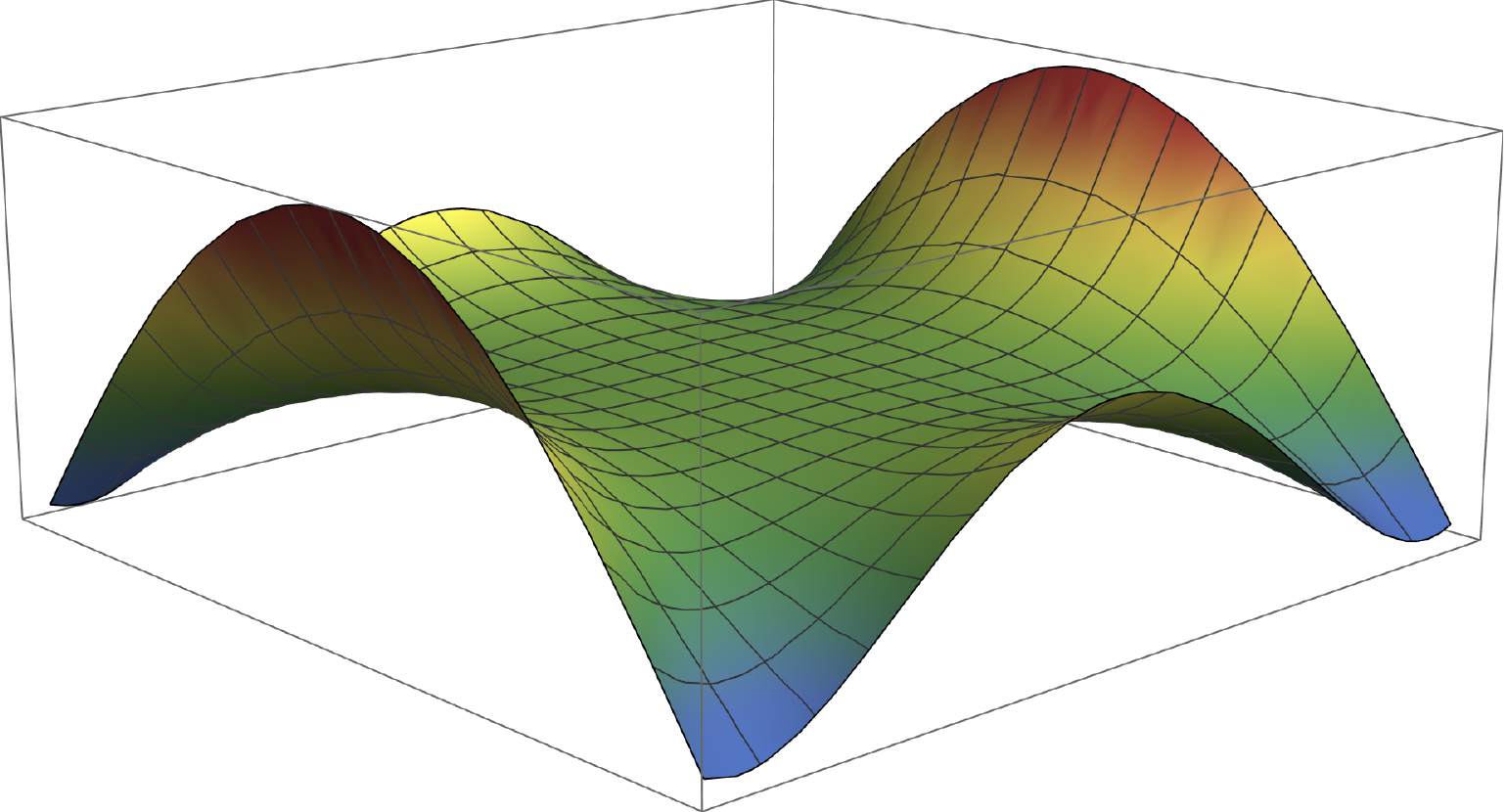}
         \caption{$\alpha=\frac{\pi}{3}$}
         \label{fig:three sin x}
     \end{subfigure}
     \hfill
     \begin{subfigure}[b]{0.3\textwidth}
         \centering
         \includegraphics[width=\textwidth]{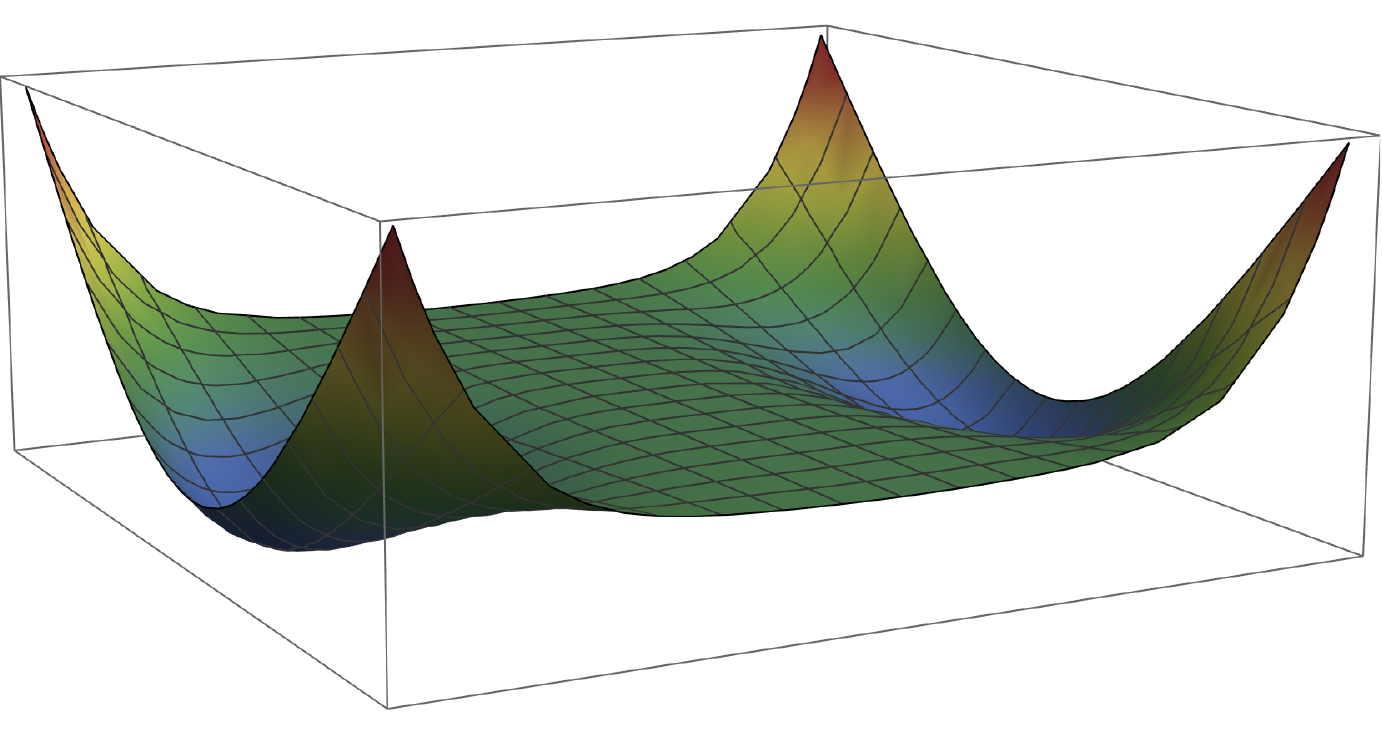}
         \caption{$\alpha=\frac{\pi}{20}$}
         \label{fig:five over x}
     \end{subfigure}
        \caption{The one-loop effective potential. Here we choose $\lambda_3=-0.9$ and $\mu = 1000$ for better visualization.}
        \label{fig1loop}
\end{figure}

\begin{figure}[h]  
\centering          
\includegraphics[width=10cm]{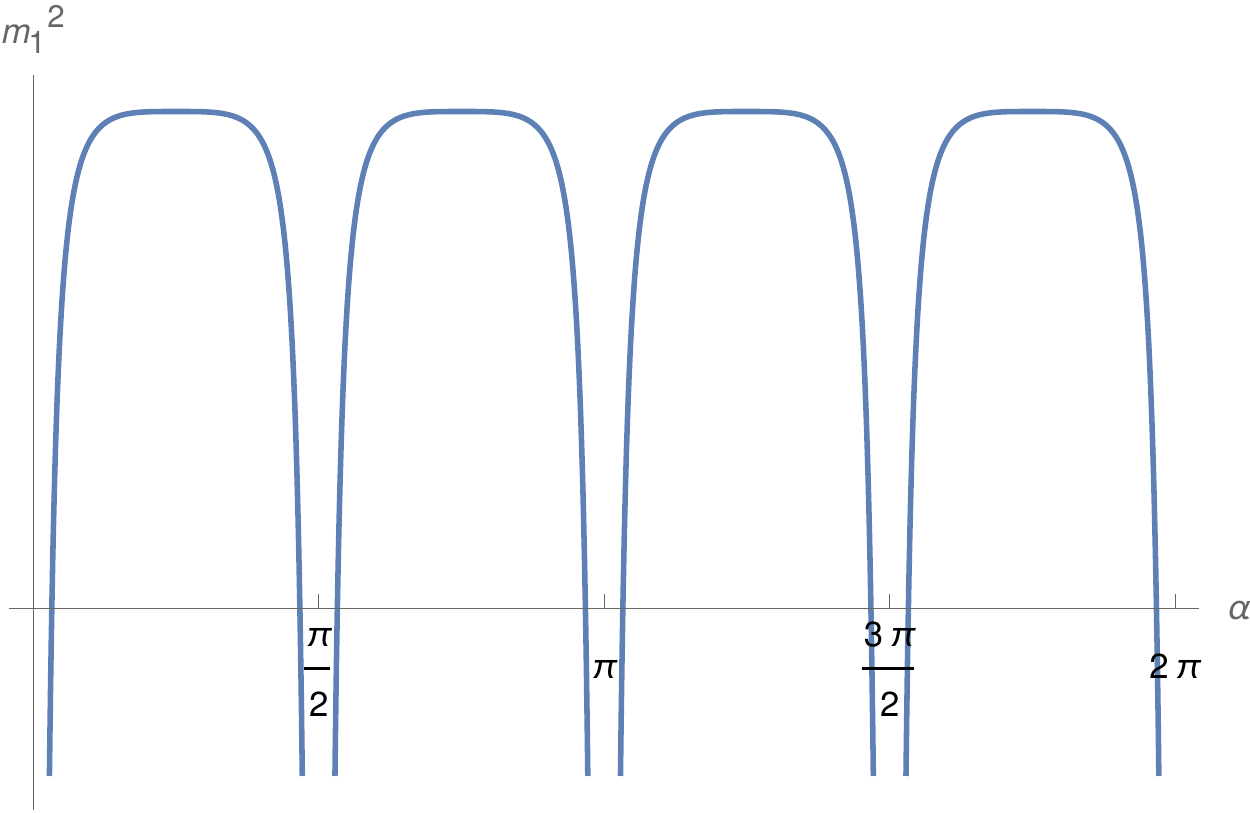}          
\caption{The dependence of mass on $\alpha$ for the one-loop effective potential.}\label{mass_alpha} \end{figure}

As demonstrated, the radiative corrections have produced a nonzero minimum in the effective potential. This can be observed graphically in Fig.~\ref{fig1loop} for certain values of $\alpha$. With this information, we determine the value of $V_1^{LL}$ at the minimum $(\sigma,\phi)=(\mu\sin\alpha,\mu\cos\alpha)$ as
\begin{equation}\label{Vmin_1loop}
 V^{LL}_1\Big{|}_{\substack{\sigma=v\sin\alpha\\ \phi=v\cos\alpha}} = -\frac{\lambda_3^2 \mu ^4}{128 \pi ^2},
\end{equation}
\noindent that is independent of $\alpha$.

It is noteworthy that the one-loop renormalization group functions enable us to calculate the complete one-loop effective potential, that is, to order $L$. In the following section, we will present the LL corrections to the effective potential up to order $L^4$.  

\subsection{Leading log corrections}\label{LeadingLogs}

The most significant outcomes of our investigation are already evident at the second logarithmic order ($L^2$). For the purpose of clarity, we will thoroughly compute the effective potential at this order and present only the results of the fourth logarithmic order potential ($L^4$) in the following. 

The LL effective potential is described by Eq.\eqref{LL}, where the coefficient $C_1^{LL}$ is given in Eq.\eqref{c1LL} and the coefficient $C_2^{LL}$ is obtained through the recurrence relation specified in Eq.\eqref{recurrence}. The expression for $C_2^{LL}$ can be cast as
\begin{eqnarray}
 C_2^{LL} &=& \frac{1}{8192 \pi ^4}\Bigr[\sigma ^4 \left(3 \lambda_1^3+\lambda_3^2 (4 \lambda_1+\lambda_2)+4 \lambda_3^3\right)+2 \lambda_3 \sigma ^2 \phi ^2 (2 \lambda_1^2+6 \lambda_3 (\lambda_1+\lambda_2)\nonumber\\
  &&+\lambda_1 \lambda_2+2 \lambda_2^2+19 \lambda_3^2)+\phi ^4 \left(\lambda_3^2 (\lambda_1+4 \lambda_2)+3 \lambda_2^3+4 \lambda_3^3\right)\Bigr].
\end{eqnarray}

Using the renormalization conditions in Eqs. \eqref{ren_conds} and the minima conditions in Eqs. \eqref{minimum}, we find the relations between the coupling constants as
\begin{subequations}\label{2loops_relations}
 \begin{eqnarray}
  &&\lambda_1 = -3 \lambda_3 \cot ^2(\alpha )-\frac{\lambda_3^2 (-37 \cos (2 \alpha )+284 \cos (4 \alpha )+29 \cos (6 \alpha )+68 \cos (8 \alpha )+88) \csc ^4(\alpha )}{128 \pi ^2}\nonumber\\
  &&\qquad -\frac{\lambda_3^3\csc ^6(\alpha ) \sec ^2(\alpha )}{131072 \pi ^4}\Bigr(-83145 \cos (2 \alpha )+528680 \cos (4 \alpha )-91591 \cos (6 \alpha )+326432 \cos (8 \alpha )\nonumber\\
  &&\qquad+6499 \cos (10 \alpha )+117004 \cos (12 \alpha )+5353 \cos (14 \alpha )+13728 \cos (16 \alpha )+68 \cos (18 \alpha )\nonumber\\
  &&\qquad+204 \cos (20 \alpha )+393280\Bigr);\\
  &&\lambda_2 = -3 \lambda_3 \tan ^2(\alpha ) -\frac{\lambda_3^2 (37 \cos (2 \alpha )+284 \cos (4 \alpha )-29 \cos (6 \alpha )+68 \cos (8 \alpha )+88) \sec ^4(\alpha )}{128 \pi ^2} \nonumber\\
  &&\qquad -\frac{\lambda_3^3\csc ^2(\alpha ) \sec ^6(\alpha )}{131072 \pi ^4}\Bigr(83145 \cos (2 \alpha )+528680 \cos (4 \alpha )+91591 \cos (6 \alpha )+326432 \cos (8 \alpha )\nonumber\\
  &&\qquad-6499 \cos (10 \alpha )+117004 \cos (12 \alpha )-5353 \cos (14 \alpha )+13728 \cos (16 \alpha )-68 \cos (18 \alpha )\nonumber\\
  &&\qquad+204 \cos (20 \alpha )+393280\Bigr).
 \end{eqnarray}
\end{subequations}

The values of $\lambda_1$ and $\lambda_2$ that minimize the effective potential are utilized to express the effective potential. By using these solutions, the LL effective potential is obtained as
\begin{eqnarray} \label{veffL2} 
V_2^{LL} &=& C_0(\lambda_2,\alpha,\sigma,\phi)+C_1(\lambda_2,\alpha,\sigma,\phi)\ln \left(\frac{\sigma ^2+\phi ^2}{\mu ^2}\right)+C_2(\lambda_2,\alpha,\sigma,\phi)\ln^2\left(\frac{\sigma ^2+\phi ^2}{\mu ^2}\right),
\end{eqnarray}
\noindent where the coefficients $C_i$ are functions of the couplings and fields, and their expressions, Eqs.~\eqref{coef0}, \eqref{coef1} and \eqref{coef2}, can be found in the Appendix \ref{appendix-coefficients}.

The mass matrix is defined by Eq.~\eqref{massmatrix}, with its components specified in Eq.\eqref{components-mass-L2} in the Appendix \ref{appendix-coefficients}. The eigenvalues of this matrix correspond to the dynamically generated masses, which are given by
\begin{subequations}
 \begin{eqnarray}
  m_{1}^2 &=& |\lambda_3| \mu ^2-\frac{|\lambda_3|^2 \mu ^2 \Big(76 \cos (4 \alpha )+17 (\cos (8 \alpha )+3)\Big) \csc ^2(2 \alpha )}{32 \pi ^2} + \frac{|\lambda_3|^3 \mu ^2\csc ^4(2\alpha )}{8192 \pi ^4} \Big(172304 \cos (4 \alpha )\nonumber\\
  &&+92772 \cos (8 \alpha )+28541 \cos (12 \alpha )+3364 \cos (16 \alpha )+51 \cos (20 \alpha )+108472\Big);\\
  m_{2}^2 &=& \frac{|\lambda_3|^2 \mu ^2}{8 \pi ^2}+\frac{|\lambda_3|^3 \mu ^2 (475 \cos (4 \alpha )+202 \cos (8 \alpha )+17 \cos (12 \alpha )+146) \csc ^2(\alpha )}{512 \pi ^4}.
 \end{eqnarray}
\end{subequations}

\begin{figure}[h]
     \centering
     \begin{subfigure}[b]{0.3\textwidth}
         \centering
         \includegraphics[width=6cm]{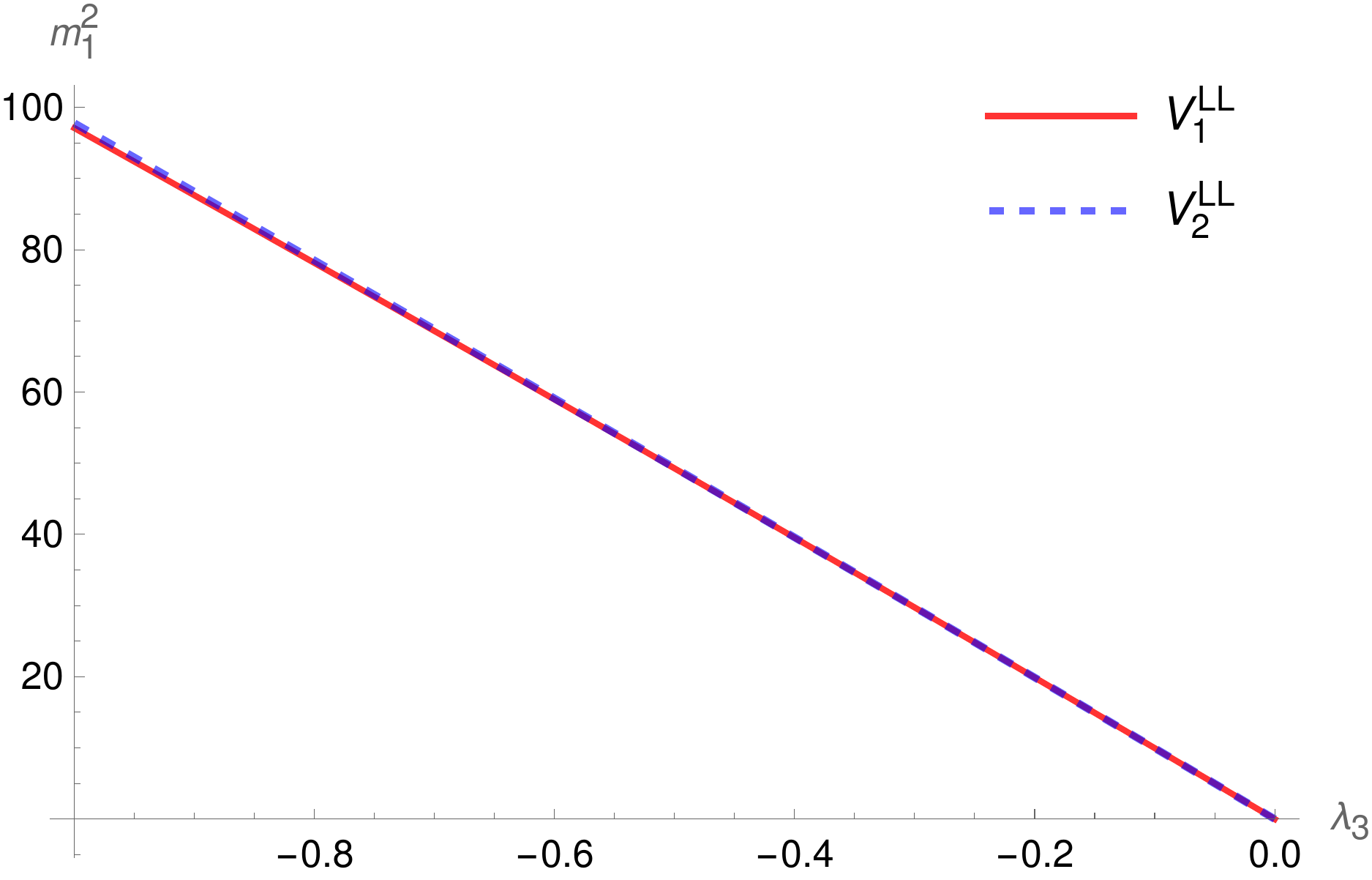}
     \end{subfigure}\hspace{3cm}
     \begin{subfigure}[b]{0.3\textwidth}
         \centering
         \includegraphics[width=6cm]{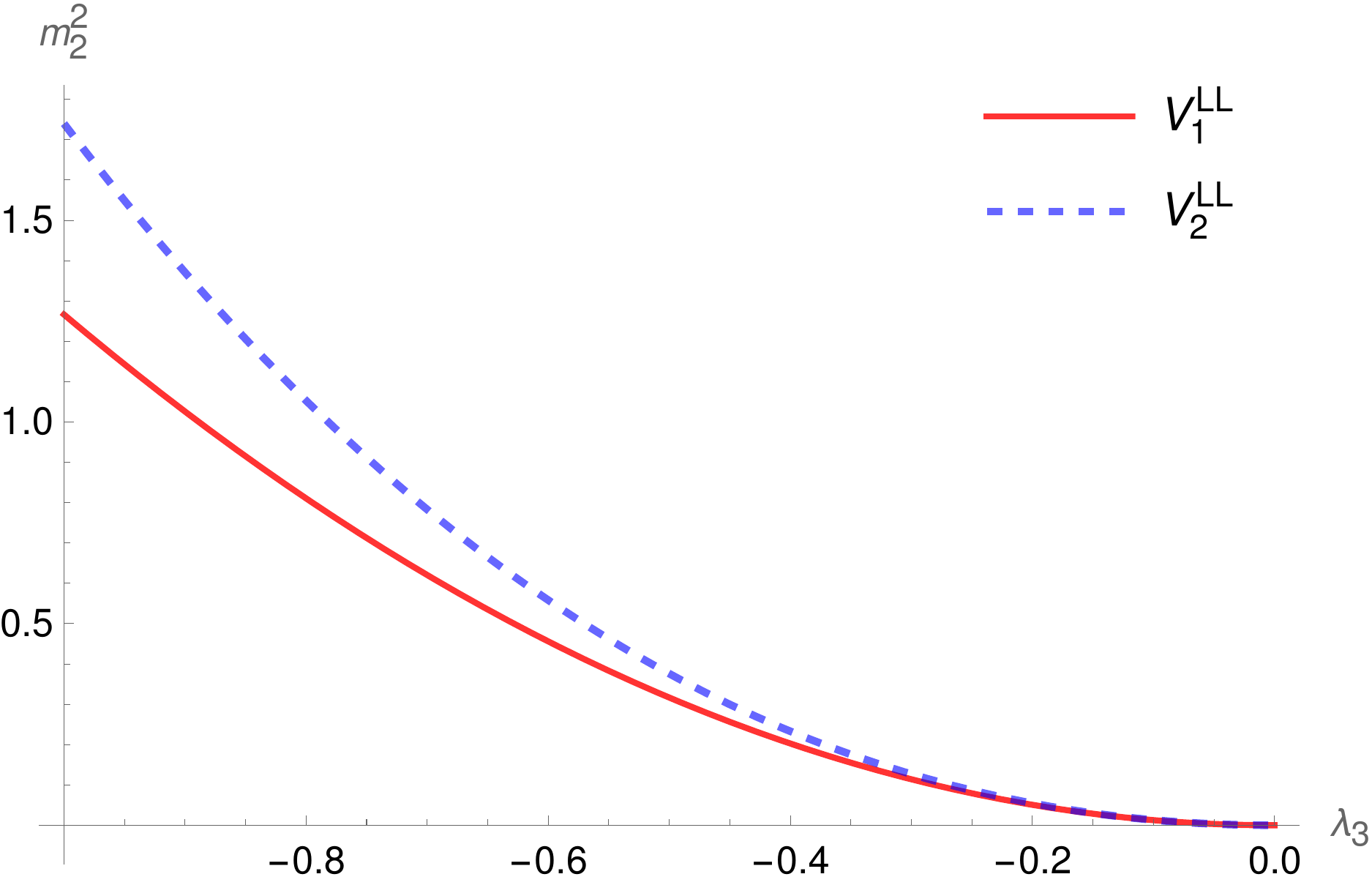}
         \label{fig:three sin x}
     \end{subfigure}
     \caption{Here we present a comparison between the masses obtained from the 1-loop and $L^2$ corrected potentials, denoted as $m_1^2$ and $m_2^2$, respectively. The parameters selected for this comparison are $\mu=10$GeV and $\alpha = 0.50948$.}
     \label{masses_comparison}
\end{figure}

As shown in Fig.~\ref{masses_comparison}, the correction to the Higgs boson is minimal, whereas the correction to the scalon is significant. For $\lambda_3=0.5$, the correction to $m_1^2$ is approximately $0.17\%$, while the correction to $m_2^2$ is $19\%$.

The evaluation of the effective potential at the minimum reveals an intriguing characteristic of the model. At the $L^2$ order, the presence of a favored flat direction for the symmetry breaking is observed. $V^{LL}_2$ evaluated at the minimum is given by
 \begin{equation}\label{Vmin_2loop}
 V^{LL}_{2}\Big{|}_{\substack{\sigma=\mu\sin\alpha\\ \phi=\mu\cos\alpha}} = -\frac{\lambda_3^2 \mu ^4}{128 \pi ^2}-\frac{\lambda_3^3 \mu ^4 (479 \cos (4 \alpha )+202 \cos (8 \alpha )+17 \cos (12 \alpha )+166) \csc ^2(\alpha ) \sec ^2(\alpha )}{32768 \pi ^4}.
\end{equation}

\begin{figure}[h]
 \centering
 \includegraphics[width=10cm]{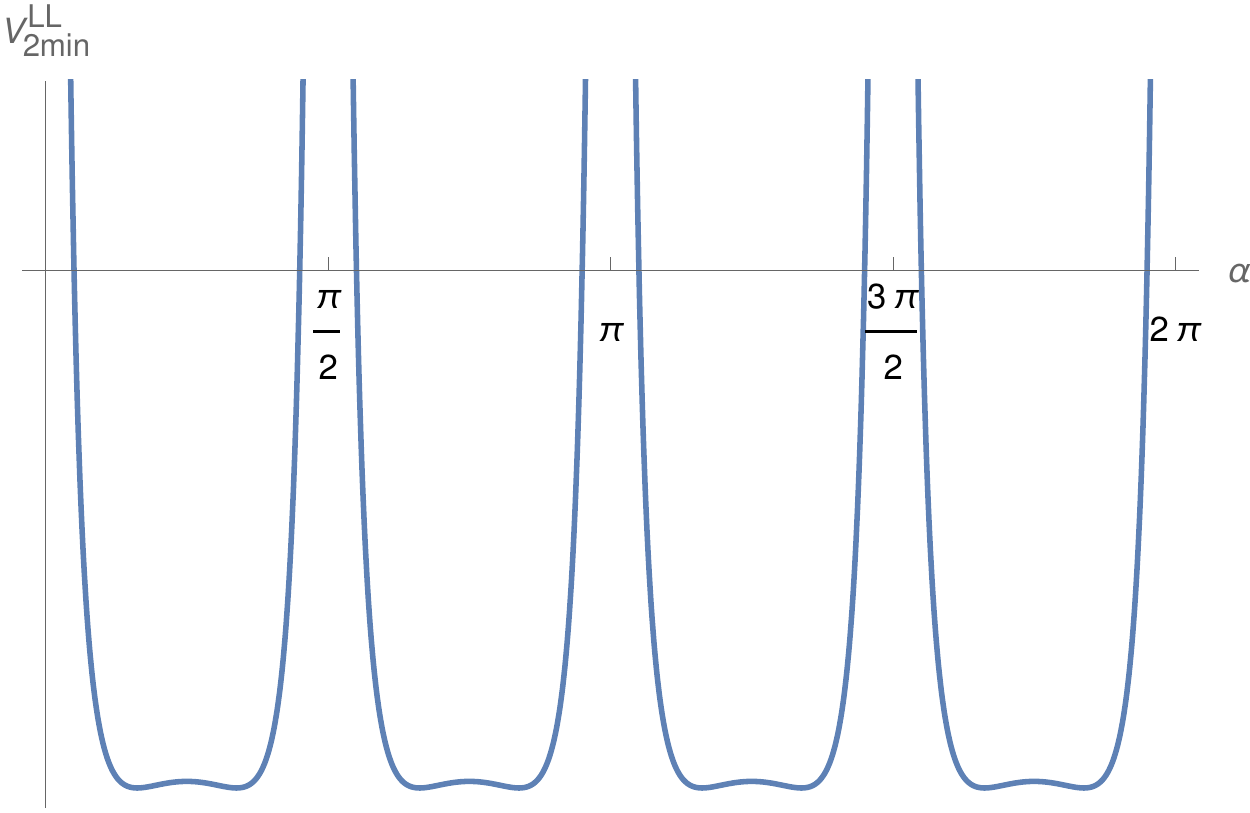}
 \caption{The behavior of $V_{2min}^{LL}$ as a function of $\alpha$}
 \label{Vmin_behavior}
\end{figure}

As evidenced, the minimum of the effective potential has a dependency on $\alpha$. Thus, it is possible to find a value of $\alpha$ that minimizes Eq.\eqref{Vmin_2loop} by analyzing its behavior as a function of $\alpha$ (as illustrated in Fig.~\ref{Vmin_behavior}). There are eight equivalent minima which can be computed to obtain numerical values that are independent of both $\lambda_3$ and $\mu$,
\begin{eqnarray*}
 &&\alpha_1 =0.50948,\quad \alpha_2 =1.06132,\quad \alpha_3 =2.08028,\quad \alpha_4 =2.63211,\nonumber\\
 && \alpha_5 =3.65107,\quad \alpha_6 =4.20291,\quad \alpha_7 =5.22187,\quad \alpha_8 =5.77370.
\end{eqnarray*}

It is possible to compute the LL corrections to any order of $L$; however, as is characteristic of the LL approximation, the corrections to the masses of the fields only appear up to $L^4$. This is due to the CW renormalization conditions as stated in Eq.~\eqref{ren_conds}. Therefore, we shall restrict our computation of these corrections to fourth order in $L$.


To compute the fourth-order LL effective potential, we apply Eq.\eqref{LL} and perform the analysis up to $n=4$. The coefficients $C_n^{LL}$ are computed using the recurrence relation in Eq.\ref{recurrence} and are given by
\begin{subequations}
 \begin{eqnarray}
  &&C_3^{LL} = \frac{1}{786432 \pi ^6}\Bigr(\phi ^4 \left(\lambda_3^2 \left(5 \lambda_1^2+10 \lambda_1 \lambda_2+47 \lambda_2^2\right)+4 \lambda_3^3 (5 \lambda_1+11 \lambda_2)+27 \lambda_2^4+63 \lambda_3^4\right)\nonumber\\
  &&\qquad\quad+\sigma ^4 \left(27 \lambda_1^4+\lambda_3^2 \left(47 \lambda_1^2+10 \lambda_1 \lambda_2+5 \lambda_2^2\right)+4 \lambda_3^3 (11 \lambda_1+5 \lambda_2)+63 \lambda_3^4\right)\nonumber\\
  &&\qquad\quad+4 \lambda_3 \sigma ^2 \phi ^2 (7 \lambda_1^3+\lambda_3 \left(19 \lambda_1^2+14 \lambda_1 \lambda_2+19 \lambda_2^2\right)+3 \lambda_1^2 \lambda_2+3 \lambda_1 \lambda_2^2+60 \lambda_3^2 (\lambda_1+\lambda_2)\nonumber\\
  &&\qquad\quad+7 \lambda_2^3+132 \lambda_3^3)\Bigr);\\
  &&C_4^{LL} = \frac{1}{25165824 \pi ^8}\Bigr(\phi ^4 (4 \lambda_3^3 \left(10 \lambda_1^2+17 \lambda_1 \lambda_2+40 \lambda_2^2\right)+\lambda_3^2 (10 \lambda_1^3+15 \lambda_1^2 \lambda_2+36 \lambda_1 \lambda_2^2\nonumber\\
  &&\qquad\quad+175 \lambda_2^3)+3 \lambda_3^4 (46 \lambda_1+91 \lambda_2)+81 \lambda_2^5+300 \lambda_3^5)+\sigma ^4 (81 \lambda_1^5+4 \lambda_3^3 (40 \lambda_1^2+17 \lambda_1 \lambda_2\nonumber\\
  &&\qquad\quad+10 \lambda_2^2)+\lambda_3^2 (175 \lambda_1^3+36 \lambda_1^2 \lambda_2+15 \lambda_1 \lambda_2^2+10 \lambda_2^3)+3 \lambda_3^4 (91 \lambda_1+46 \lambda_2)+300 \lambda_3^5)\nonumber\\
  &&\qquad\quad+2 \lambda_3 \sigma ^2 \phi ^2 (35 \lambda_1^4+14 \lambda_1^3 \lambda_2+2 \lambda_3^2 \left(146 \lambda_1^2+127 \lambda_1 \lambda_2+146 \lambda_2^2\right)\nonumber\\
  &&\qquad\quad+30 \lambda_3 (\lambda_1+\lambda_2) \left(3 \lambda_1^2-\lambda_1 \lambda_2+3 \lambda_2^2\right)+12 \lambda_1^2 \lambda_2^2+14 \lambda_1 \lambda_2^3+702 \lambda_3^3 (\lambda_1+\lambda_2)\nonumber\\
  &&\qquad\quad+35 \lambda_2^4+1236 \lambda_3^4).
 \end{eqnarray}
\end{subequations}

By employing a similar approach as previously, we can determine the value of the fourth-order LL effective potential at the minimum as
\begin{eqnarray}
 V^{LL}_{4}\Big{|}_{\substack{\sigma=\mu\sin\alpha\\ \phi=\mu\cos\alpha}}&=&-\frac{\lambda_3^2 \mu ^4}{128 \pi ^2}-\frac{\lambda_3^3 \mu ^4 \csc ^2(2 \alpha )}{8192 \pi ^4}(479 \cos (4 \alpha )+202 \cos (8 \alpha )+17 \cos (12 \alpha )+166)\nonumber\\
 &&-\frac{\lambda_3^4 \mu ^4\csc ^4(2 \alpha )}{8388608 \pi ^6}\Bigr(4603564 \cos (4 \alpha )+3038264 \cos (8 \alpha )+1494988 \cos (12 \alpha )\nonumber\\
 && +304245 \cos (16 \alpha ) +19848 \cos (20 \alpha )+204 \cos (24 \alpha )+3256967\Bigr)\nonumber\\
 &&
 -\frac{\lambda_3^5 \mu ^4\csc ^6(2 \alpha )}{2147483648 \pi ^8}\Bigr(17713055869 \cos (4 \alpha ) +12357068468 \cos (8 \alpha )\nonumber\\
 &&+6543984311 \cos (12 \alpha )
  +2928279032 \cos (16 \alpha )+826644101 \cos (20 \alpha )\nonumber\\
 &&+116005644 \cos (24 \alpha )+6830051 \cos (28 \alpha )+125232 \cos (32 \alpha )\nonumber\\
 && +612 \cos (36 \alpha ) +9194780440\Bigr),
\end{eqnarray}

\noindent where the values of $\alpha$ that minimizes it are
\begin{eqnarray*}
 &&\alpha_1 =0.50596,\quad \alpha_2 =1.06483,\quad \alpha_3 =2.07676,\quad \alpha_4 =2.63563,\nonumber\\
 && \alpha_5 =3.64756,\quad \alpha_6 =4.20642,\quad \alpha_7 =5.21836,\quad \alpha_8 =5.77722.
\end{eqnarray*}

\begin{figure}[h]
     \centering
     \begin{subfigure}[b]{0.3\textwidth}
         \centering
         \includegraphics[width=6cm]{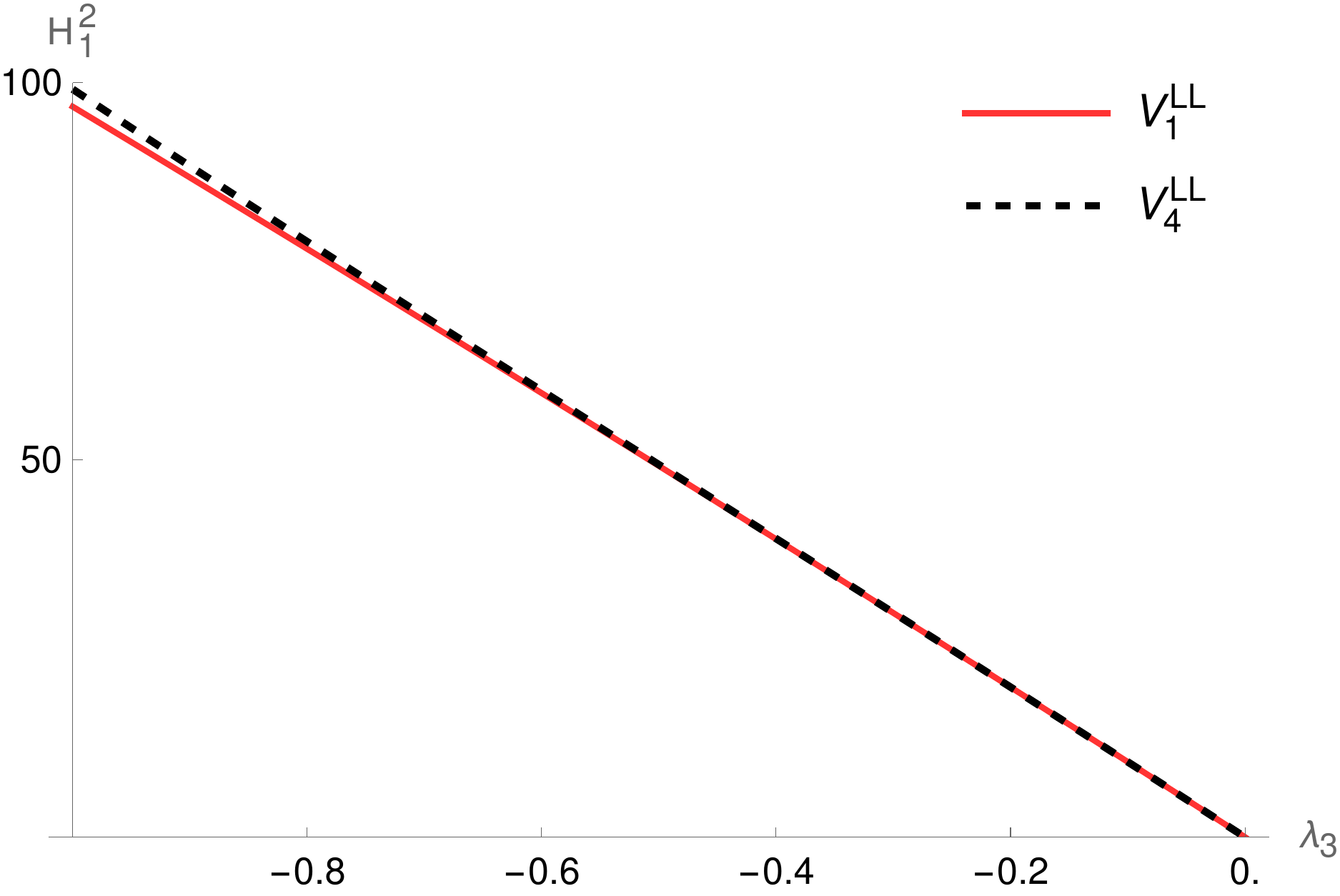}
        \caption{}
     \end{subfigure}\hspace{3cm}
     \begin{subfigure}[b]{0.3\textwidth}
         \centering
         \includegraphics[width=6cm]{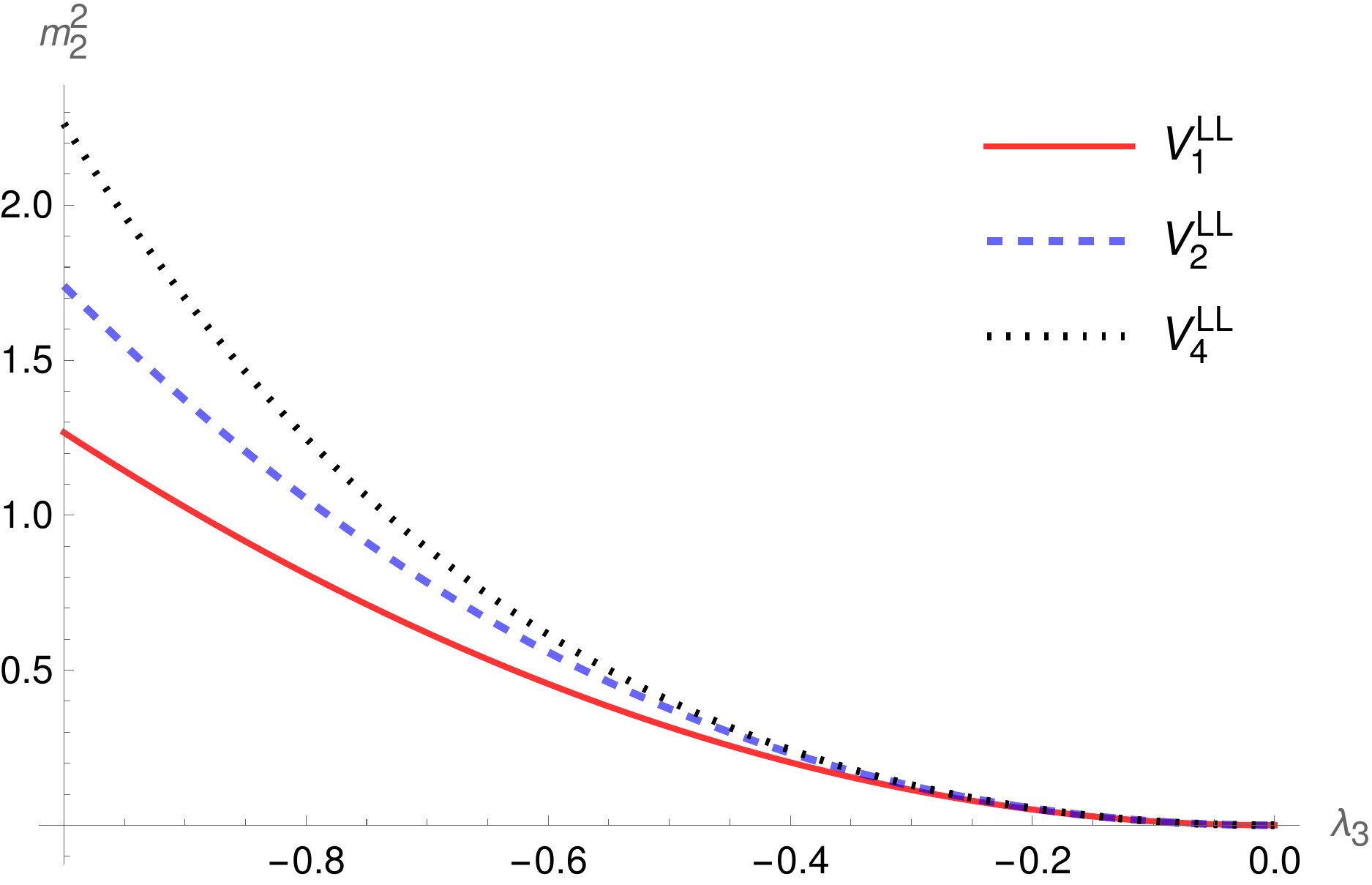}
         \caption{}
     \end{subfigure}
    \caption{(a) The Higgs mass is obtained from the expression of $V_1^{LL}$ and $V_4^{LL}$. In this case, we have selected a value of $\mu=10$GeV and $\alpha=0.505967$ as the inputs. (b) The scalon mass is calculated from the expression of $V_1^{LL}$, $V_2^{LL}$, and $V_4^{LL}$. A fixed value of $\mu=10$GeV is used in this calculation. The value of $\alpha$ is set to 0.50948 for $V_2^{LL}$ and 0.505967 for $V_4^{LL}$.}\label{masses_comparison_4loops}
\end{figure}

The correction to $\alpha$ in comparison to the $V^{LL}_{2}$ result is relatively small. Figure \ref{masses_comparison_4loops} illustrates the comparison of the masses generated by the $L^4$ calculation. The correction for the Higgs boson mass is minimal, approximately $0.33\%$ for $\lambda_3=-0.5$. However, the correction for the scalon is substantial, approximately $27\%$ when compared to the one-loop mass and approximately $6.9\%$ when compared to the $V^{LL}_{2}$ generated mass.

{Although it is possible to numerically solve the RG equation and obtain the full series of the LL effective potential, we limit our analysis to fourth order since higher orders do not contribute to the generated masses. Additionally, contributions from higher powers than $L^4$ do not affect the value of the effective potential at its minimum, as the renormalization conditions in Eqs.~\eqref{1loop-ct} are fourth derivatives evaluated at $\sigma^2+\phi^2=\mu^2$. However, it should be noted that our result will be corrected by next-to-leading Logs terms (NLL) or beyond. To compute the NLL terms, we would need to use the two-loop RG equation, and for the NNLL contributions, we would need to use the three-loop RG equation, and so on. Further details and a schematic computation are provided in Appendix \ref{Robustness}.}

\section{Final remarks}\label{Conclusions}

In this study, we evaluated the improved Gildener-Weinberg effective potential for a two-scalar interacting model. Our results demonstrate that higher-order perturbative corrections to the effective potential can significantly impact the dynamics of symmetry breaking. This is because higher-order corrections make the minimum of the effective potential dependent on the flat direction of spontaneous symmetry breaking at tree level, leading to a preferred direction for the true minimum. {It is worth noting that our approach to improving the effective potential is based on the summation of leading logarithms, which should not be confused with the RG improvement presented in Ref.\cite{Ookane:2019iwq}. In that work, the author considers $\mu^2 = \mu_0^2e^{2t}$ and expresses the renormalized coupling constants as functions of $t$. Our approach, on the other hand, involves using the RG equation to derive a recurrence relation that enables us to calculate the leading higher-loop contributions to the effective potential, as originally described in Ref.\cite{McKeon:1998tr}. This has led us to obtain a new result for the preferred flat direction compared to Ref.~\cite{Ookane:2019iwq}. A promising avenue for future research would be to combine our method with the one used in \cite{Ookane:2019iwq} to achieve a better approximation of the effective potential.}

This methodology can be applied to more general models and the outcomes can be highly informative. For instance, the Lee-Pilaftsis model, which is an extension of the Standard Model with two Higgs scalars, was studied in detail in Refs.~\cite{Lee:2012jn,Lane:2018ycs,Lane:2019dbc}. In \cite{Lane:2018ycs}, the authors examine the potential experimental implications and detectability of the model in the LHC. In \cite{Lane:2019dbc}, the authors propose that the well-known Higgs boson is actually the scalon from the Gildener-Weinberg model, while the heavier particle is a new Higgs boson with a mass of up to 550 GeV. This demonstrates the significance of higher-loop analyses, as the correction to the scalon mass can be substantial. Additionally, the preferred direction of the effective potential should be treated with caution. In previous works \cite{Lee:2012jn,Lane:2018ycs,Lane:2019dbc}, the authors had the freedom to choose the angle $\alpha$ as they saw fit. However, our higher-order corrections highlight the importance of a more careful treatment of this aspect.

Another model that our approach could be applied to is the one proposed in Ref. \cite{Held:2022hnw}. In this work, the authors present a grand unification effective field theory in which the symmetry breaking occurs through the Gildener-Weinberg mechanism. The authors thoroughly study the possible and viable symmetry breaking patterns.

It should be noted that special attention may be required when applying this method to gauge theories. The effective potential in gauge theories can be gauge dependent, as demonstrated in Refs.~\cite{Jackiw:1974cv,Nielsen:1975fs}. Therefore, care should be taken to include gauge-dependent artifacts such as daisies, as discussed in Refs.~\cite{Bazeia:1988pz,deLima:1989yf,Andreassen:2014eha}. We expect to study this feature in a forthcoming paper.

Finally, one of the most intriguing applications is to Higgs portal models for dark matter \cite{Cosme:2018nly, Arcadi:2019lka, Arcadi:2021mag, Steele:2013fka}. In such models, DM is represented by a scalar field that couples to the Higgs boson in a manner similar to what we have studied in our model. The electroweak symmetry breaking may occur in various ways, with one of them being the Gildener-Weinberg mechanism as discussed in \cite{Steele:2013fka}. A crucial aspect of DM models is the abundance of DM \cite{Gondolo:1990dk}, which is sensitive to the properties of the DM field. Hence, corrections to the masses and coupling constants play a significant role in such scenarios.

\section*{ACKNOWLEDGMENTS}
The work of H.~S. and L.~H.~S.~R. is partially supported
by Coordena\c{c}\~ao de Aperfei\c{c}oamento de Pessoal de N\'ivel Superior (CAPES).

\appendix
\section{Renormalization group functions}\label{apxA}

In this appendix, we aim to calculate the renormalization group functions of the model in order to obtain the one-loop contributions to the effective potential.

\begin{figure}[h]
	\begin{center}
	\includegraphics[angle=0 ,width=10cm]{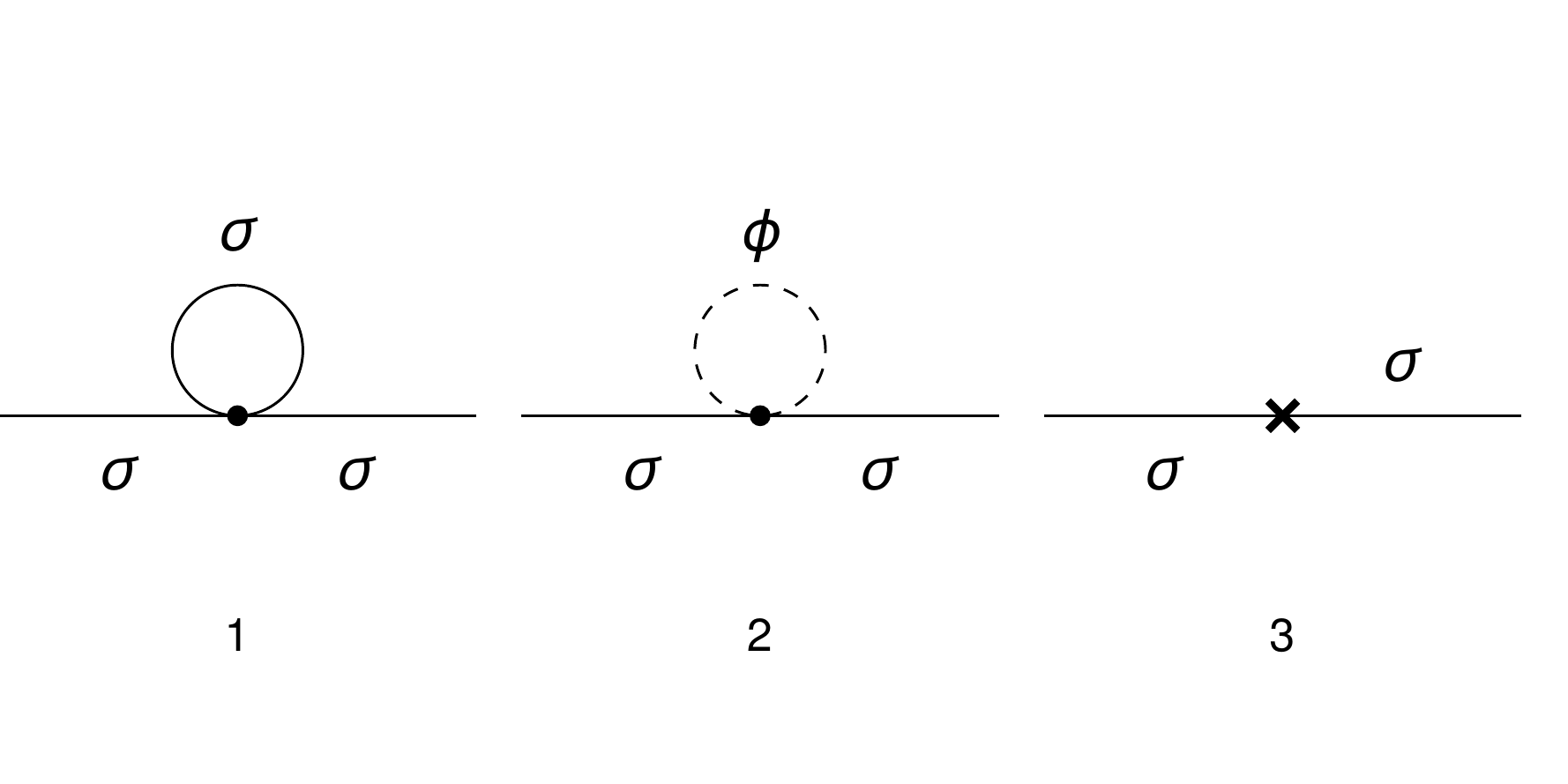}
	\caption{Feynman diagrams for the $\sigma$ field self-energy.}	\label{fig01}
	\end{center}
\end{figure}

\begin{figure}[h]
	\begin{center}
	\includegraphics[angle=0 ,width=10cm]{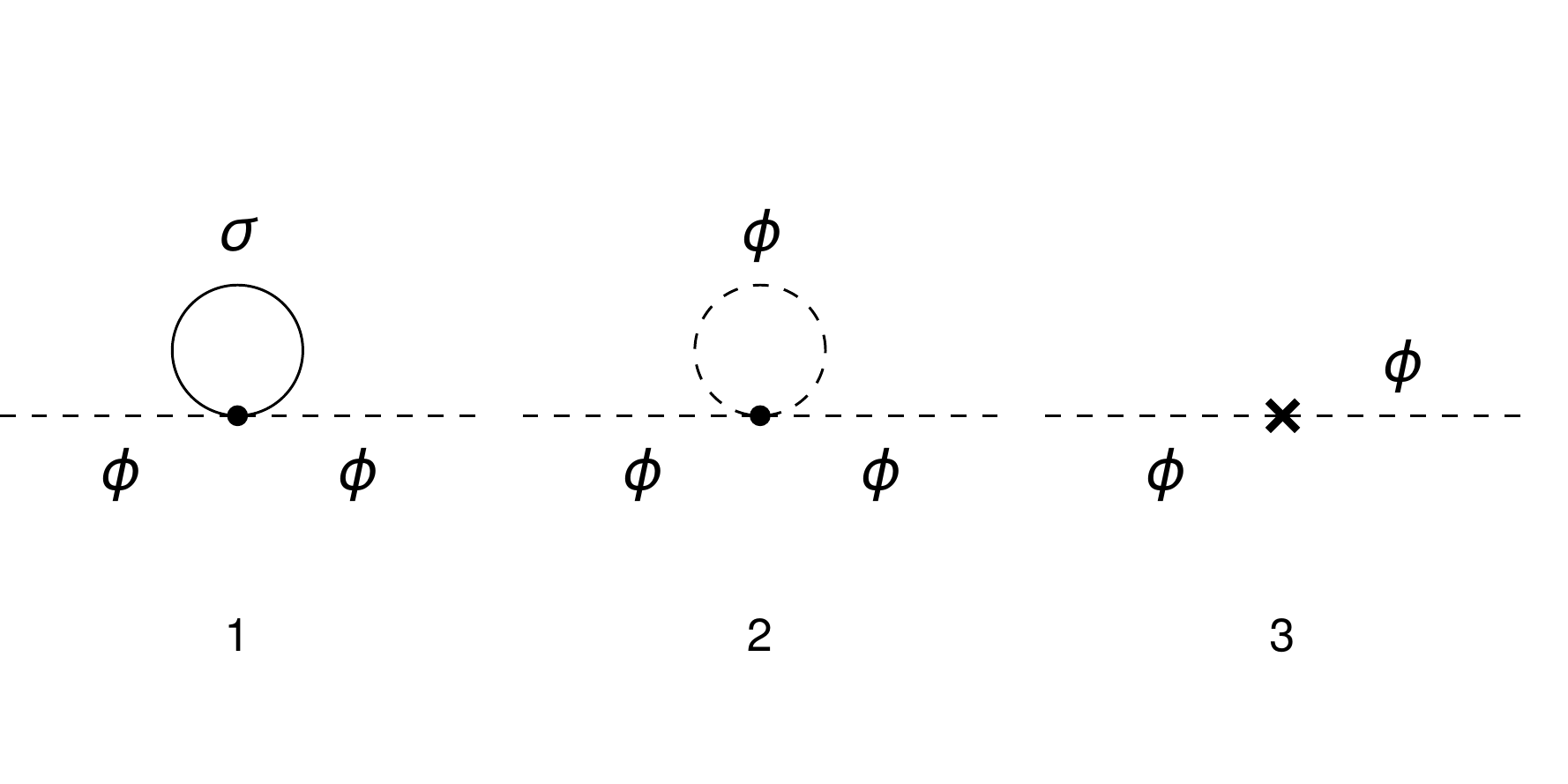}
	\caption{Feynman diagrams for the $\phi$ field self-energy.}\label{fig02}
	\end{center}
\end{figure}

To start, we evaluate the one-loop self-energies and four-point functions. We first calculate the self-energy of the $\sigma$ field. The diagrams contributing to this calculation are depicted in Fig. \ref{fig01}. The contribution is given by
\begin{equation}
\Gamma_\sigma = \int\frac{d^4k}{(2\pi)^4}\left(\frac{\lambda_1+\lambda_3}{2k^2}\right) + ip^2\delta_\sigma,
\end{equation}

\noindent where $\delta_\sigma$ represents the wave-function counterterm for the $\sigma$ field. At one-loop order, $\delta_\sigma$ is equal to zero. The same situation occurs for the self-energy of the $\phi$ field, as shown in Fig. \ref{fig02}. In this case, the wave-function counterterm, $\delta_\phi$, is equal to zero as well.

\begin{figure}[h]
	\begin{center}
	\includegraphics[angle=0 ,width=10cm]{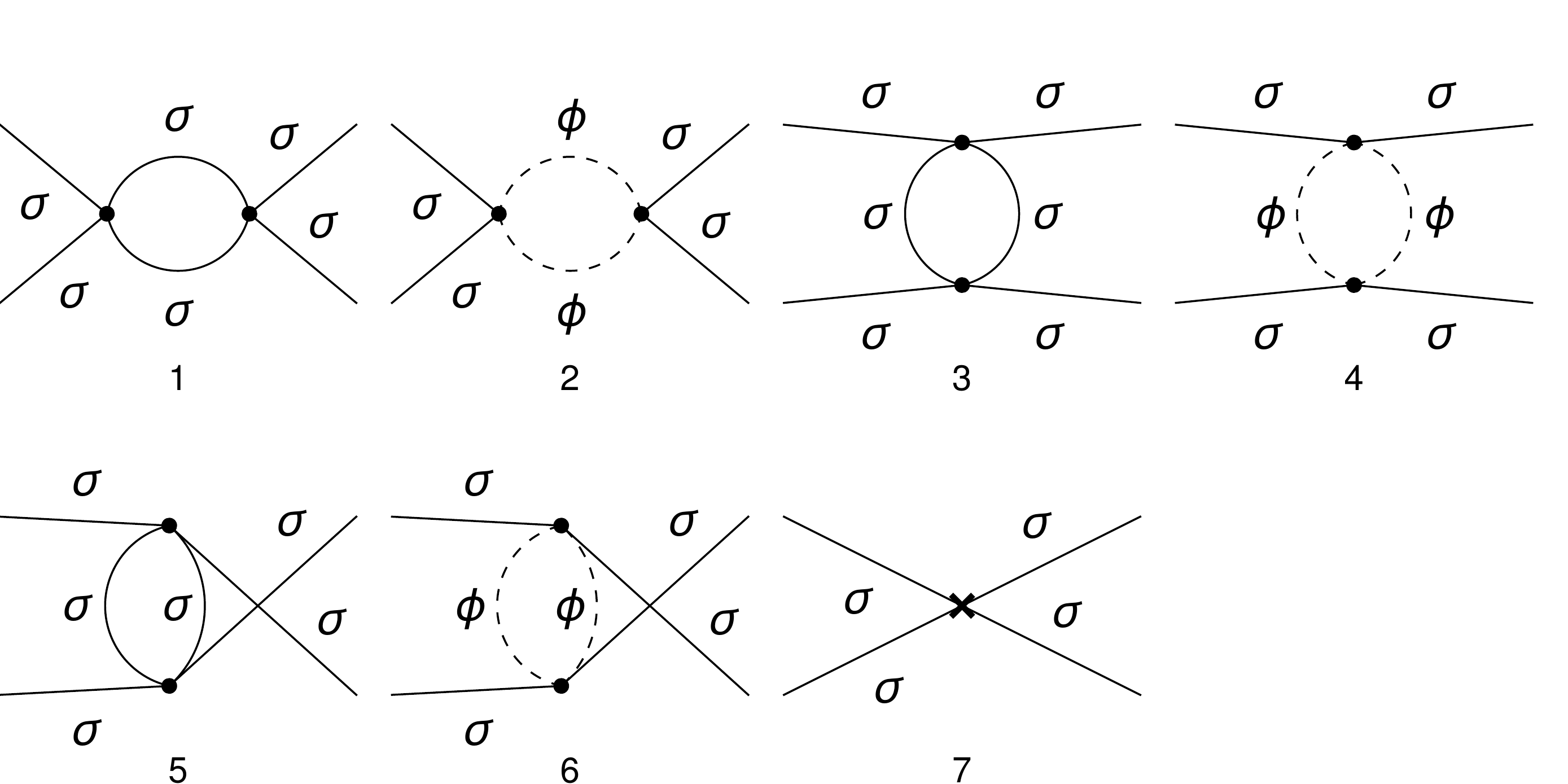}
	\caption{Feynman diagrams for the $\sigma^4$ vertex function.}\label{fig03}
	\end{center}
\end{figure}

Next, we evaluate the four-point functions. The first calculation involves the four-point function of the $\sigma$ field, which is depicted in Fig.~\ref{fig03}. The contribution from these diagrams is given by
\begin{equation}
\Gamma_2 = \frac{(\lambda_1^2 + \lambda_3^2)}{2}\int\frac{d^4k}{(2\pi)^4}\left(\frac{1}{k^2(k+p_2-p_3)^2} + \frac{1}{k^2(k+p_2-p_4)^2} + \frac{1}{k^2(k-p_3-p_4)^2}\right) - i\lambda_1\delta_{\lambda_1},
\end{equation}
where $p_1$ and $p_2$ represent incoming momenta, and $p_3$ and $p_4$ represent outgoing momenta. In order to impose finiteness through minimal subtraction (MS), we have
\begin{equation}
\delta_{\lambda_1} = \frac{3(\lambda_1^2 + \lambda_3^2)}{32\pi^2\lambda_1\epsilon}.
\end{equation}

\begin{figure}[h]
	\begin{center}
	\includegraphics[angle=0 ,width=10cm]{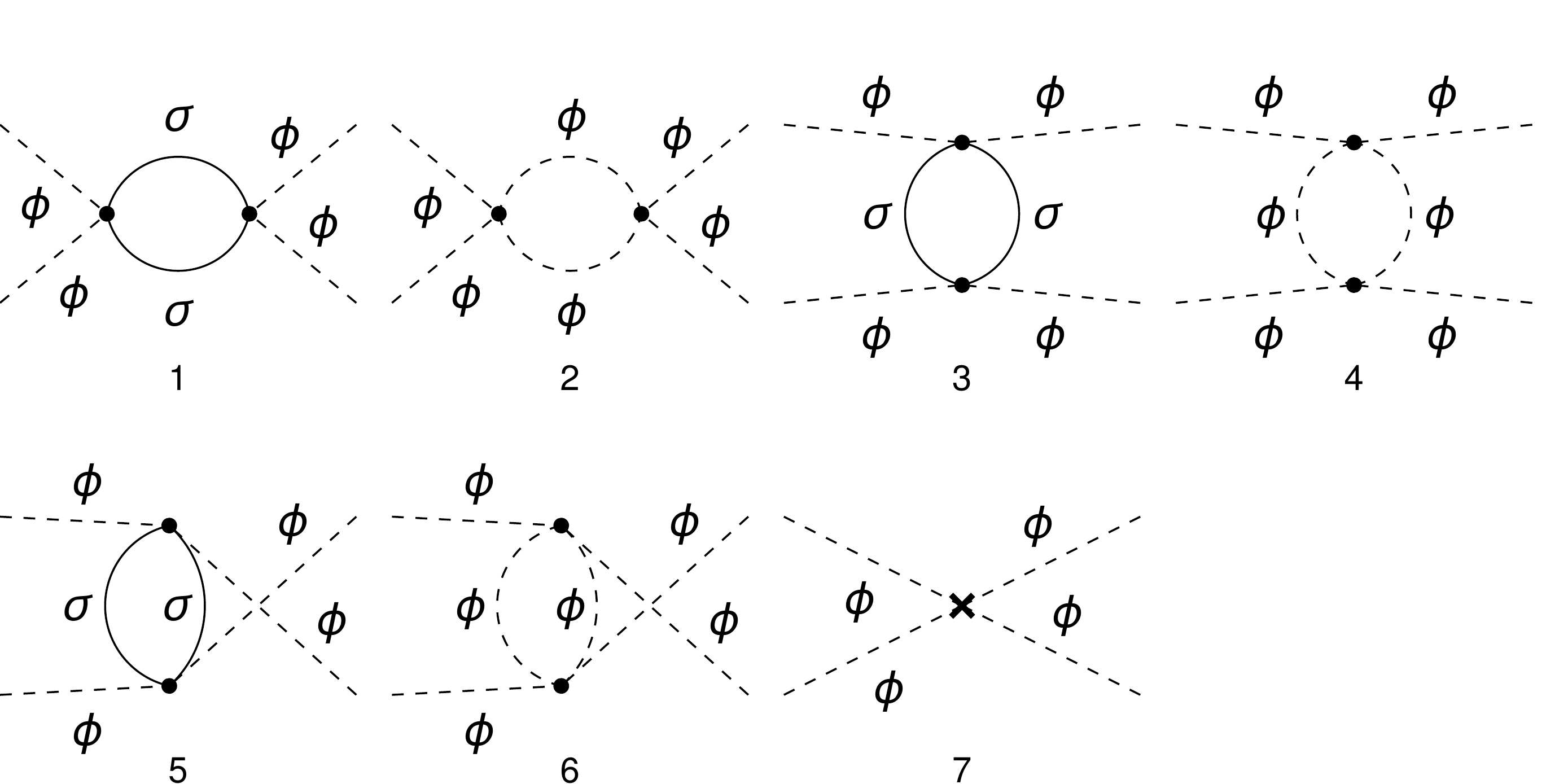}
	\caption{Feynman diagrams for the $\phi^4$ vertex function.}\label{fig04}
	\end{center}
\end{figure}

In a similar way, the diagrams in Fig. \ref{fig04} give the following contribution,
\begin{equation}
 \Gamma_3 = \frac{(\lambda_2^2 + \lambda_3^2)}{2}\int\frac{d^4k}{(2\pi)^4}\left(\frac{1}{k^2(k+p_2-p_3)^2} + \frac{1}{k^2(k+p_2-p_4)^2} + \frac{1}{k^2(k-p_3-p_4)^2}\right) - i\lambda_2\delta_{\lambda_2},
\end{equation}
therefore,
\begin{equation}
 \delta_{\lambda_2} = \frac{3(\lambda_2^2 + \lambda_3^2)}{32\pi^2\lambda_2\epsilon}
\end{equation}

\begin{figure}[h]
	\begin{center}
	\includegraphics[angle=0 ,width=10cm]{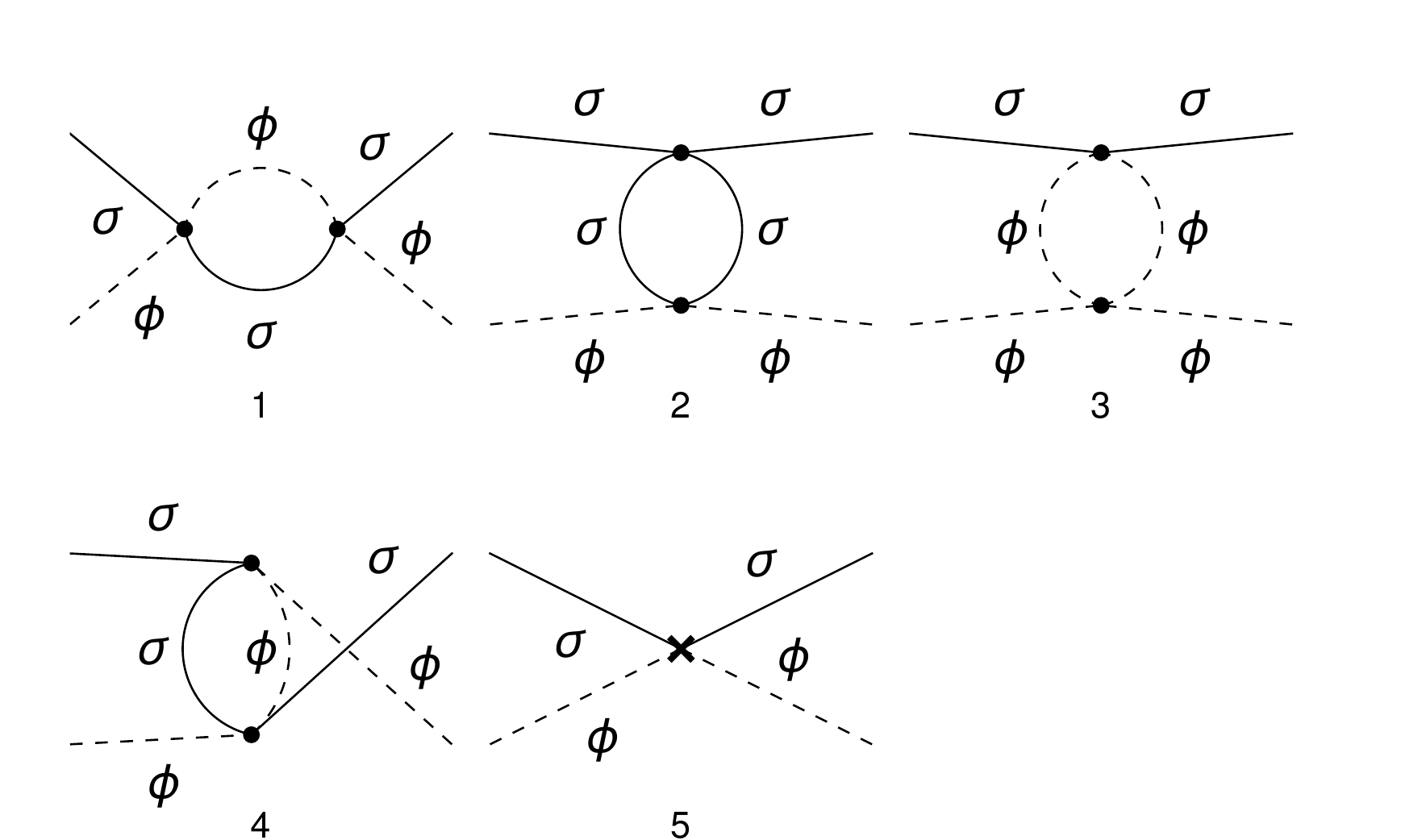}
	\caption{Feynman diagrams for the $\sigma^2\phi^2$ vertex function.}\label{fig05}
	\end{center}
\end{figure}

And finally, for the diagrams in Fig. \ref{fig05} we have
\begin{equation}
 \Gamma_4 = \lambda_3\int\frac{d^4k}{(2\pi)^4}\left(\frac{(\lambda_1+\lambda_2)}{2k^2(k+p_2-p_4)} + \frac{\lambda_3}{k^2(k+p_2-p_4)^2} + \frac{\lambda_3}{k^2(k-p_3-p_4)^2}\right) - i\lambda_3\delta_{\lambda_3},
\end{equation}
\noindent where imposing finiteness through MS we find
\begin{equation}
 \delta_{\lambda_3} = \frac{(\lambda_1+\lambda_2+4\lambda_3)}{32\pi^2\epsilon}.
\end{equation}

Finally, we have the following renormalization group functions \cite{Ookane:2019iwq},
\begin{equation}
 \beta_{\lambda_1}=\lim_{\epsilon\rightarrow0}\mu\frac{\partial\lambda_1}{\partial\mu}=\frac{3}{16\pi^2}(\lambda_1^2+\lambda_3^2);
\end{equation}
\begin{equation}
 \beta_{\lambda_2}=\lim_{\epsilon\rightarrow0}\mu\frac{\partial\lambda_2}{\partial\mu}=\frac{3}{16\pi^2}(\lambda_2^2+\lambda_3^2);
\end{equation}
\begin{equation}
 \beta_{\lambda_3}=\lim_{\epsilon\rightarrow0}\mu\frac{\partial\lambda_3}{\partial\mu}=\frac{\lambda_3}{16\pi^2}(\lambda_1+\lambda_2+4\lambda_3);
\end{equation}
\begin{equation}
 \gamma_\sigma=\gamma_\phi=0.
\end{equation}

\section{Effective potential with trivial $\alpha$}\label{apxB}

Now, let us consider the scenario in which $\alpha=\frac{n\pi}{2}$, with $n\in\mathbb{Z}$. In this case, either $\sin\alpha=0$ or $\cos\alpha=0$. The conditions set forth in Eqs.~\eqref{tadpoles} state that
\begin{eqnarray}
\langle\eta_1(0)\rangle &=& -i\left(\frac{\lambda_1}{6}\sin^2\alpha + \frac{\lambda_3}{2}\cos^2\alpha\right)\mu\sin\alpha = 0, \\
\langle\eta_2(0)\rangle &=& -i\left(\frac{\lambda_2}{6}\cos^2\alpha + \frac{\lambda_3}{2}\sin^2\alpha\right)\mu\cos\alpha = 0.
\end{eqnarray}

\noindent This implies that one of the self-interaction coupling constants must be zero. For example, if we consider the case where $\alpha=0$, then either $\mu=0$ or $\lambda_2=0$. Assuming that the coupling constants cannot be zero, we conclude that $\mu=0$, meaning that at tree level, there is no symmetry breaking or mass generation.

We can move to the one-loop analyses with the ansatz
\begin{equation}
 V_{eff} = A + BL,
\end{equation}
in which $L = \ln\left(\frac{\sigma^2+\phi^2}{\mu^2}\right)$. In this case, after using the renormalization conditions in Eqs. \eqref{ren_conds} we obtain
\begin{eqnarray}
 V_1 &=& \sigma ^4 \left(\frac{-4 \lambda_3 (\lambda_1+\text{$\lambda $2})+\text{$\lambda $2}^2-15 \lambda_3^2}{512 \pi ^2}+\frac{\lambda_1}{24}\right)-\frac{\sigma ^2 \phi ^2 \left(\lambda_3 \left(6 \lambda_1+25 \lambda_3-64 \pi ^2\right)+\text{$\lambda $2}^2+6 \text{$\lambda $2} \lambda_3\right)}{256 \pi ^2}\nonumber\\
 &&+\phi ^4 \left(\frac{\text{$\lambda $2}}{24}-\frac{25 \left(\text{$\lambda $2}^2+\lambda_3^2\right)}{1536 \pi ^2}\right)+\frac{1}{256 \pi ^2} \Big[\sigma ^4 \left(\lambda_1^2+\lambda_3^2\right)+2 \lambda_3 \sigma ^2 \phi ^2 (\lambda_1+\text{$\lambda $2}+4 \lambda_3)\nonumber\\
 &&+\phi ^4 \left(\text{$\lambda $2}^2+\lambda_3^2\right)\Big]\ln \left(\frac{\sigma ^2+\phi ^2}{\mu ^2}\right).
\end{eqnarray}

The minimum conditions for the potential can be written
\begin{subequations}  \begin{eqnarray}   &&\qquad\qquad\qquad\frac{\partial V_1}{\partial \sigma}\Bigr |_{\substack{\sigma=0\\\phi=\mu}} =0;\\   &&\frac{\partial V_1}{\partial \phi}\Bigr |_{\substack{\sigma=0\\\phi=\mu}} =-\frac{11 \text{$\lambda $2}^2 \mu ^3}{192 \pi ^2}+\frac{\text{$\lambda $2} \mu ^3}{6}-\frac{11 \lambda_3^2 \mu ^3}{192 \pi ^2}, \end{eqnarray} \end{subequations}
\noindent where these conditions lead to the following relationship between the coupling constants $\lambda_2$ and $\lambda_3$:
\begin{equation}
\lambda_2 = \frac{11 \lambda_3^2}{32 \pi ^2}.
\end{equation}

Therefore, the effective potential becomes
\begin{eqnarray}
 V_1 &=& \sigma ^4 \left(\frac{\lambda_1}{24}-\frac{\lambda_3 (4 \lambda_1+15 \lambda_3)}{512 \pi ^2}\right)+\frac{\lambda_3 \sigma ^2 \phi ^2 \left(-6 \lambda_1-25 \lambda_3+64 \pi ^2\right)}{256 \pi ^2}-\frac{\lambda_3^2 \phi ^4}{512 \pi ^2} \nonumber\\
 &&+ \frac{1}{256 \pi ^2}\left(\sigma ^4 \left(\lambda_1^2+\lambda_3^2\right)+2 \lambda_3 \sigma ^2 \phi ^2 (\lambda_1+4 \lambda_3)+\lambda_3^2 \phi ^4\right) \ln \left(\frac{\sigma ^2+\phi ^2}{\mu ^2}\right).
\end{eqnarray}

The mass matrix derived from $V_1$ is diagonal and its eigenvalues are given by 
\begin{subequations}
\begin{eqnarray}
m^2_\sigma &=& \frac{\lambda_3\mu^2}{2} - \frac{3\lambda_1\lambda_3\mu^2}{64\pi^2} - \frac{3\lambda_3^2\mu^2}{16\pi^2}\\
m^2_\phi &=& \frac{\lambda_3^2\mu^2}{32\pi^2}. 
\end{eqnarray}
\end{subequations}

\noindent So, the minimum of $V_1$ is located at $(\sigma,\phi) = (0,\mu)$ for perturbative couplings that satisfy the conditions $\lambda_3 > 0$ and $\lambda_1 < 4(8\pi^2 - 3\lambda_3)/3$.

The value of $V_1$ at the minimum of the potential can be expressed as
\begin{eqnarray}
V_{1}\Big{|}_{\substack{\sigma=0\\ \phi=\mu}} = -\frac{\lambda_3^2 \mu ^4}{512 \pi ^2}.
\end{eqnarray}

\noindent From Eq.\eqref{Vmin_1loop}, it can be seen that the minimum of the potential is deeper for the case in which $\alpha \neq \frac{n\pi}{2}, n\in\mathbb{Z}$, making this case more favorable.

\section{Large logarithms}\label{LargeLogs}
The purpose of this appendix is to explain why large logarithms are not present in our analysis. To begin, let us consider the problem in the context of the Coleman-Weinberg mechanism \cite{Coleman:1973jx}. As demonstrated in the original work by Coleman and Weinberg, the effective potential for the $\lambda\phi^4$ model can be given by
\begin{equation}
 V_{eff} = \frac{\lambda}{4!}\phi^4 + \frac{\lambda^2\phi^4}{256\pi^2}\left(\log\frac{\phi^2}{\mu^2} - \frac{25}{6}\right).
\end{equation}

Thus, we observe that by imposing the minimum condition 
\begin{equation}\label{minimum}
 \frac{d V_{eff}}{d\phi}\Bigr|_{\phi = \langle\phi\rangle} = 0,
\end{equation}
we arrive at the expression
\begin{equation}
 \lambda\log\frac{\langle\phi\rangle^2}{\mu^2} = -\frac{32\pi^2}{3} + O(\lambda).
\end{equation}
\noindent  where $\mu$ is the renormalization scale.


The effective potential suffers from large logarithms due to the quantity $\lambda\log\frac{\langle\phi\rangle}{\mu^2}$ being outside the perturbative region. This issue arises because only one coupling constant is present in the $\lambda\phi^4$ model, which is not the case in scalar electrodynamics. The effective potential for this model is given by
\begin{equation}
V_{eff} = \frac{\lambda}{4!}\phi^4 + \left(\frac{5\lambda^2}{1152\pi^2} + \frac{3e^4}{64\pi^2}\right)\phi^4\log\frac{\phi^2}{\mu^2}.
\end{equation}

Using the minimum condition, we can derive the expression
\begin{equation}
\frac{\left(54 e^4+5 \lambda ^2\right) \left(3 \log \frac{\langle\phi\rangle^2}{\mu ^2}-11\right)}{864 \pi ^2}+\frac{\lambda}{6} = 0.
\end{equation}
We can rewrite this as
\begin{equation}
\log\frac{\langle\phi^2\rangle}{\mu^2} = \frac{11}{3} - \frac{144\pi^2\lambda}{3(54e^4+5\lambda^2)}.
\end{equation}
We can then find the minimum as
\begin{equation}
\langle\phi\rangle^2 = \mu^2\exp\left(\frac{11}{3} - \frac{144\pi^2\lambda}{3(54e^4+5\lambda^2)}\right).
\end{equation}

As $\mu$ is an arbitrary renormalization scale, we can choose it to be the most convenient, as discussed in Ref.~\cite{Coleman:1973jx}. In this case, we choose $\mu = \langle\phi\rangle$ and solve the above equation perturbatively to find
\begin{equation}
\lambda = \frac{33e^4}{8}.
\end{equation}

Henceforth, the effective potential can be expressed as
\begin{equation}
V_{eff} = \frac{3e^4}{64\pi^2}\phi^4\left(\log\frac{\phi^2}{\langle\phi\rangle^2} -\frac{1}{2}\right),
\end{equation}
and there is no issue with large logarithms as long as we confine ourselves to $\phi$ values close to the minimum. The same applies to the two-scalar model discussed in this study. By following the same method as described above, we obtain the coupling constant relationships presented in Eqs.~\eqref{couplingRelations}. Therefore, we can conclude that the effective potential is applicable for values of $\sigma^2+\phi^2$ that are sufficiently near to the minimum, such that $\lambda_3^2L$ is within the perturbative range.

\section{The $L^2$ effective potential coefficients}\label{appendix-coefficients}

In this appendix, we present the coefficients that appear in the expression for the $L^2$ effective potential, as defined in Eq.~\eqref{veffL2}. Additionally, we provide detailed expressions for the corresponding mass matrix.

The coefficients appearing in the Eq.\eqref{veffL2} are given by
\begin{eqnarray}\label{coef0}
C_0(\lambda_3,\alpha,\sigma,\phi)&=& \frac{\lambda_3}{8} \left(\sigma ^4 \left(-\csc ^2(\alpha )\right)-\phi ^4 \sec ^2(\alpha )+\left(\sigma ^2+\phi ^2\right)^2\right) -\frac{\lambda_3^2}{128 \pi ^2} \Big[136 \cos (2 \alpha ) \left(\sigma ^4-\phi ^4\right) \nonumber\\
&& +34 \cos (4 \alpha ) \left(\sigma ^2+\phi ^2\right)^2+18 \Big(\sigma ^4 \csc ^4(\alpha )-2 \sigma ^2 \csc ^2(\alpha ) \left(4 \sigma ^2+\phi ^2\right)  \nonumber\\
&& -2 \phi ^2 \sec ^2(\alpha ) \left(\sigma ^2+4 \phi ^2\right)+\phi ^4 \sec ^4(\alpha )\Big)+247 \sigma ^4+222 \sigma ^2 \phi ^2+247 \phi ^4\Big]  \nonumber\\
&& +\frac{\lambda_3^3}{4096 \pi ^4} \Big[ 213864 \cos (2 \alpha ) \left(\sigma ^4-\phi ^4\right)+14884 \cos (6 \alpha ) \left(\sigma ^4-\phi ^4\right) \nonumber\\
&& +204 \cos (10 \alpha ) \left(\sigma ^4-\phi ^4\right)-1584 \sigma ^4 \csc ^6(\alpha )+51 \cos (12 \alpha ) \left(\sigma ^2+\phi ^2\right)^2 \nonumber\\
&& +36 \sigma ^2 \csc ^4(\alpha ) \left(699 \sigma ^2+88 \phi ^2\right)+36 \phi ^2 \sec ^4(\alpha ) \left(88 \sigma ^2+699 \phi ^2\right) \nonumber\\
&& +\cos (4 \alpha ) \left(72339 \sigma ^4+85958 \sigma ^2 \phi ^2+72339 \phi ^4\right)+2 \cos (8 \alpha ) \left(2005 \sigma ^4+3602 \sigma ^2 \phi ^2+2005 \phi ^4\right) \nonumber\\
&& -4 \csc ^2(\alpha ) \left(42172 \sigma ^4+11844 \sigma ^2 \phi ^2+423 \phi ^4\right)-4 \sec ^2(\alpha ) \left(423 \sigma ^4+11844 \sigma ^2 \phi ^2+42172 \phi ^4\right) \nonumber\\
&& -1584 \phi ^4 \sec ^6(\alpha )+4 \left(80247 \sigma ^4+60902 \sigma ^2 \phi ^2+80247 \phi ^4\right)\Big],
\end{eqnarray}
\begin{eqnarray}\label{coef1}
C_1(\lambda_3,\alpha,\sigma,\phi)&=& 
\frac{\lambda_3^2}{256 \pi ^2} \Big[ 9 \sigma ^4 \csc ^4(\alpha )-6 \sigma ^2 \csc ^2(\alpha ) \left(3 \sigma ^2+\phi ^2\right)-6 \phi ^2 \sec ^2(\alpha ) \left(\sigma ^2+3 \phi ^2\right)+9 \phi ^4 \sec ^4(\alpha ) \nonumber\\
&&+10 \left(\sigma ^2+\phi ^2\right)^2\Big] +\frac{\lambda_3^3}{1024 \pi ^4} \Big[ -1719 \cos (2 \alpha ) \left(\sigma ^4-\phi ^4\right)-27 \sigma ^2 \csc ^4(\alpha ) \left(37 \sigma ^2+\phi ^2\right)\nonumber\\
&&+81 \sigma ^4 \csc ^6(\alpha )+9 \sigma ^2 \csc ^2(\alpha ) \left(405 \sigma ^2+34 \phi ^2\right)
-27 \phi ^2 \sec ^4(\alpha ) \left(\sigma ^2+37 \phi ^2\right)\nonumber\\
&& +9 \phi ^2 \sec ^2(\alpha ) \left(34 \sigma ^2+405 \phi ^2\right)-68 \cos (4 \alpha ) \left(3 \sigma ^4+2 \sigma ^2 \phi ^2+3 \phi ^4\right)\nonumber\\
&&+81 \phi ^4 \sec ^6(\alpha )-6 \left(707 \sigma ^4+180 \sigma ^2 \phi ^2+707 \phi ^4\right)\Big],
\end{eqnarray}
\noindent and 
\begin{eqnarray}\label{coef2}
C_2(\lambda_3,\alpha,\sigma,\phi)&=& \frac{\lambda_3^3 }{8192 \pi ^4} \Big[ -81 \sigma ^4 \csc ^6(\alpha )+9 \sigma ^2 \csc ^4(\alpha ) \left(27 \sigma ^2+4 \phi ^2\right)+9 \phi ^2 \sec ^4(\alpha ) \left(4 \sigma ^2+27 \phi ^2\right)\nonumber\\
&&-3 \csc ^2(\alpha ) \left(85 \sigma ^4+36 \sigma ^2 \phi ^2+\phi ^4\right)-3 \sec ^2(\alpha ) \left(\sigma ^4+36 \sigma ^2 \phi ^2+85 \phi ^4\right)\nonumber\\
&&-81 \phi ^4 \sec ^6(\alpha )+100 \left(\sigma ^2+\phi ^2\right)^2\Big].
\end{eqnarray}

The expressions for the components of the mass matrix can be cast as
\begin{subequations}\label{components-mass-L2}
 \begin{eqnarray}
  &&\frac{\partial^2 V^{LL}}{\partial \sigma^2}\Bigr |_{\substack{\sigma=\mu\sin\alpha\\\phi=\mu\cos\alpha}} = -\lambda_3 \mu ^2 \cos ^2(\alpha )-\frac{\lambda_3^2 \mu ^2 (-16 \cos (2 \alpha )+86 \cos (4 \alpha )+17 \cos (8 \alpha )+57) \csc ^2(\alpha )}{128 \pi ^2}\nonumber\\
  &&\qquad\qquad\qquad\qquad-\frac{\lambda_3^3 \mu ^2 \csc ^4(\alpha ) \sec ^2(\alpha )}{131072 \pi ^4} \Bigr(-21752 \cos (2 \alpha )+190946 \cos (4 \alpha )-19180 \cos (6 \alpha )\nonumber\\
  &&\qquad\qquad\qquad\qquad+104520 \cos (8 \alpha )-3612 \cos (10 \alpha )+31019 \cos (12 \alpha )+3432 \cos (16 \alpha )\nonumber\\
  &&\qquad\qquad\qquad\qquad+51 \cos (20 \alpha )+120080\Bigr);\\
  &&\frac{{\partial^2 V^{LL}}}{\partial \phi^2}\Bigr |_{\substack{\sigma=\mu\sin\alpha\\\phi=\mu\cos\alpha}} = -\lambda_3 \mu ^2 \sin ^2(\alpha )-\frac{\lambda_3^2 \mu ^2 (16 \cos (2 \alpha )+86 \cos (4 \alpha )+17 \cos (8 \alpha )+57) \sec ^2(\alpha )}{128 \pi ^2}\nonumber\\
  &&\qquad\qquad\qquad\qquad-\frac{\lambda_3^3 \mu ^2 \csc ^2(\alpha ) \sec ^4(\alpha )}{131072 \pi ^4} (21752 \cos (2 \alpha )+190946 \cos (4 \alpha )+19180 \cos (6 \alpha )\nonumber\\
  &&\qquad\qquad\qquad\qquad+104520 \cos (8 \alpha )+3612 \cos (10 \alpha )+31019 \cos (12 \alpha )+3432 \cos (16 \alpha )\nonumber\\
  &&\qquad\qquad\qquad\qquad+51 \cos (20 \alpha )+120080);\\
  && \frac{\partial^2 V^{LL}}{\partial \sigma\partial\phi}\Bigr |_{\substack{\sigma=\mu\sin\alpha\\\phi=\mu\cos\alpha}} = \lambda_3 \mu ^2 \sin (\alpha ) \cos (\alpha )+\frac{\lambda_3^2 \mu ^2 (86 \cos (4 \alpha )+17 \cos (8 \alpha )+65) \csc (2 \alpha )}{64 \pi ^2}\nonumber\\
  &&\qquad\qquad\qquad\qquad+\frac{\lambda_3^3 \mu ^2\csc ^3(\alpha ) \sec ^3(\alpha )}{131072 \pi ^4} (204994 \cos (4 \alpha )+110984 \cos (8 \alpha )+31563 \cos (12 \alpha )\nonumber\\
  &&\qquad\qquad\qquad\qquad+3432 \cos (16 \alpha )+51 \cos (20 \alpha )+123600);\\
  && \frac{\partial^2 V^{LL}}{\partial \phi\partial\sigma}\Bigr |_{\substack{\sigma=\mu\sin\alpha\\\phi=\mu\cos\alpha}} = \lambda_3 \mu ^2 \sin (\alpha ) \cos (\alpha )+\frac{\lambda_3^2 \mu ^2 (86 \cos (4 \alpha )+17 \cos (8 \alpha )+65) \csc (2 \alpha )}{64 \pi ^2}\nonumber\\
  &&\qquad\qquad\qquad\qquad+\frac{\lambda_3^3 \mu ^2\csc ^3(\alpha ) \sec ^3(\alpha )}{131072 \pi ^4} (204994 \cos (4 \alpha )+110984 \cos (8 \alpha )+31563 \cos (12 \alpha )\nonumber\\
  &&\qquad\qquad\qquad\qquad+3432 \cos (16 \alpha )+51 \cos (20 \alpha )+123600).
 \end{eqnarray}
\end{subequations}


\section{Accuracy and limitations of the $L^4$ approximation}\label{Robustness}

As discussed in Sec.~\ref{LeadingLogs}, it is possible to truncate the leading logs approximation at $L^4$ since higher powers will not improve the generated masses or the minimum of the potential. In this appendix, we demonstrate schematically that this is true due to the renormalization conditions presented in Eqs.~\eqref{ren_conds}.


Let us take the leading logs effective potential before renormalization
\begin{equation}\label{effPot}
 V^{LL} = \frac{\delta_1}{24}\sigma^4 + \frac{\delta_2}{24}\phi^4 + \frac{\delta_3}{4}\sigma^2\phi^2 + \sum_{n=0}^\infty A_nL^n
\end{equation}
in which $A_n = A_n(\sigma,\phi,\lambda_1,\lambda_2,\lambda_3)$. For convenience, we will express $A_n$ as $A_n(\sigma,\phi)$. In this way, we obtain
\begin{eqnarray}\label{D4sigma}
 \frac{\partial^4V^{LL}}{\partial\sigma^4} &=& \frac{1}{\left(\sigma ^2+\phi ^2\right)^4}\left(\sigma ^2+\phi ^2\right)^4 A_0^{(4,0)}(\sigma ,\phi )+8 \sigma  \left(\sigma ^2+\phi ^2\right)^3 A_1^{(3,0)}(\sigma ,\phi )\nonumber\\
 &&-12 \left(\sigma ^2-\phi ^2\right) \left(\sigma ^2+\phi ^2\right)^2 A_1^{(2,0)}(\sigma ,\phi )+16 \left(\sigma ^2+\phi ^2\right) \left(\sigma ^3-3 \sigma  \phi ^2\right) A_1^{(1,0)}(\sigma ,\phi )\nonumber\\
 &&-12 \left(\sigma ^4-6 \sigma ^2 \phi ^2+\phi ^4\right) A_1(\sigma ,\phi )+48 \sigma ^2 \left(\sigma ^2+\phi ^2\right)^2 A_2^{(2,0)}(\sigma ,\phi )+96 \sigma  \phi ^2 \left(\sigma ^2+\phi ^2\right) A_2^{(1,0)}(\sigma ,\phi )\nonumber\\
 &&-96 \sigma ^3 \left(\sigma ^2+\phi ^2\right) A_2^{(1,0)}(\sigma ,\phi )+88 \sigma ^4 A_2(\sigma ,\phi )-240 \sigma ^2 \phi ^2 A_2(\sigma ,\phi )+24 \phi ^4 A_2(\sigma ,\phi )\nonumber\\
 &&+192 \sigma ^3 \left(\sigma ^2+\phi ^2\right) A_3^{(1,0)}(\sigma ,\phi )-288 \left(\sigma ^4-\sigma ^2 \phi ^2\right) A_3(\sigma ,\phi )+384 \sigma ^4 A_4(\sigma ,\phi ) + O(L),
\end{eqnarray}
in which
\begin{equation}
 A_n^{(a,b)}(\sigma,\phi)=\frac{\partial^{(a+b)}A_n(\sigma,\phi)}{\partial^a\sigma\partial^b\phi}.
\end{equation}

It is observed that the $A_n$ terms only exist up to $n=4$. Therefore, upon substituting $\sigma = \mu\sin\alpha$ and $\phi = \mu\cos\alpha$ in Eq.~\eqref{D4sigma} (and similarly for the other two renormalization conditions), all the terms proportional to $O(L)$ vanish and $A_n$, with $n>4$, do not contribute to the counterterms. Hence, the renormalized effective potential can be expressed as
\begin{equation}
 V^{LL} = \tilde{A}_0(\sigma,\phi) + \sum_{n=1}^\infty A_n(\sigma,\phi)L^n,
\end{equation}
where $\tilde{A}_0$ now carries the counterterm information, and $A_n$ remains the same as in Eq.~\eqref{effPot} [as seen explicitly at one-loop order in Eq.\eqref{renormalized_1loop}]. Next, the minimum conditions can be imposed as
\begin{subequations}
 \begin{eqnarray}
  &&\frac{\partial V^{LL}}{\partial\sigma}\Bigr |_{\substack{\sigma=\mu\sin\alpha\\\phi=\mu\cos\alpha}} = 0;\\
  &&\frac{\partial V^{LL}}{\partial\phi}\Bigr |_{\substack{\sigma=\mu\sin\alpha\\\phi=\mu\cos\alpha}} = 0.
 \end{eqnarray}
\end{subequations}

Using the same argument as above, it is evident that the only terms contributing to the relations between coupling constants are $\tilde{A}_{0}$ and $A_1$. After this step, the relevant physical quantities can be computed. First, the value of the minimum of the potential can be calculated as
\begin{equation}
 V_{min} = V^{LL}\Bigr |_{\substack{\sigma=\mu\sin\alpha\\\phi=\mu\cos\alpha}}.
\end{equation}

Since we evaluate the effective potential at the point $\sigma^2+\phi^2 = \mu^2$, any terms proportional to $O(L)$ will vanish, and the only term contributing to the minimum will be the one proportional to $L^0$. We conclude that the $L^4$ approximation suffices to determine the minimum of the effective potential, and our result is robust because it does not receive any corrections from terms proportional to $L^m$, where $m>4$. Furthermore, the generated masses will not be corrected by any term beyond the $L^4$ approximation. However, it should be noted that our result can be corrected by next-to-leading Logs terms (NLL) or beyond. To compute the NLL terms, we would need to use the two-loop RG equation, and for the NNLL contributions, we would need to use the three-loop RG equation, and so on.

\newpage 

\end{document}